\documentclass[twocolumn,twocolappendix]{aastex63}
\usepackage{amsmath}

\newcommand\kms{km s$^{-1}$}

\newcommand\teff{$T_{eff}$}
\newcommand\logg{$\log g$}

\defcitealias{kounkel2018a}{Paper I}
\usepackage{listings}

\begin{document}
\title{ABYSS I: Targeting strategy for APOGEE \& BOSS young star survey in SDSS-V}

\author[0000-0002-5365-1267]{Marina Kounkel}
\affil{Department of Physics and Astronomy, Vanderbilt University, VU Station 1807, Nashville, TN 37235, USA}
\email{marina.kounkel@vanderbilt.edu}

\author[0000-0003-3769-8812]{Eleonora Zari}
\affiliation{Max-Planck-Institut f\"ur Astronomie, K\"onigstuhl 17 D-69117 Heidelberg, Germany}
\author[0000-0001-6914-7797]{Kevin Covey}
\affil{Department of Physics and Astronomy, Western Washington University, 516 High St, Bellingham, WA 98225, USA}
\author[0000-0003-0842-2374]{Andrew Tkachenko}
\affiliation{Institute of Astronomy, KU Leuven, Celestijnenlaan 200D, B-3001 Leuven, Belgium}
\author[0000-0001-8600-4798]{Carlos Román Zúñiga}
\affiliation{Universidad Nacional Aut\'onoma de M\'exico, Instituto de Astronom\'ia, AP 106,  Ensenada 22800, BC, M\'exico}
\author[0000-0002-3481-9052]{Keivan Stassun}
\affil{Department of Physics and Astronomy, Vanderbilt University, VU Station 1807, Nashville, TN 37235, USA}
\author[0000-0003-2300-8200]{Amelia M.\ Stutz}
\affiliation{Departamento de Astronom\'{i}a, Universidad de Concepci\'{o}n,Casilla 160-C, Concepci\'{o}n, Chile}
\author[0000-0003-1479-3059]{Guy Stringfellow}
\affil{Center for Astrophysics and Space Astronomy, Department of Astrophysical and Planetary Sciences, University of Colorado,Boulder,CO, 80309, USA}
\author[0000-0002-1379-4204]{Alexandre Roman-Lopes}
\affiliation{Departamento de Astronomia, Facultad de Ciencias, Universidad de La Serena.  Av. Juan Cisternas 1200, La Serena, Chile}
\author[0000-0001-9797-5661]{Jes\'us Hern\'andez}
\affiliation{Universidad Nacional Aut\'onoma de M\'exico, Instituto de Astronom\'ia, AP 106,  Ensenada 22800, BC, M\'exico}
\author[0000-0002-5855-401X]{Karla Pe\~na Ram\'irez}
\affiliation{Centro de Astronom\'ia (CITEVA), Universidad de Antofagasta, Av. Angamos 601, Antofagasta, Chile}
\author[0000-0001-7868-7031]{Amelia Bayo}
\affiliation{Instituto de F\'{\i}sica y Astronom\'{\i}a, Universidad de Valpara\'{\i}so, Chile}
\affil{European Southern Observatory, Karl-Schwarzschild-Strasse 2, 85748 Garching bei München, Germany}
\author[0000-0001-6072-9344]{Jinyoung Serena Kim}
\affil{Steward Observatory, Department of Astronomy, University of Arizona, 933 North Cherry Avenue, Tucson, AZ 85721-0065, USA}
\author[0000-0002-8849-9816]{Lyra Cao}
\affil{Department of Astronomy, The Ohio State University, Columbus, OH 43210, USA}
\author[0000-0002-0826-9261]{Scott J. Wolk}
\affil{Center for Astrophysics | Harvard \& Smithsonian 60 Garden St. Cambridge MA, 02138 USA}
\author[0000-0001-9852-1610]{Juna Kollmeier}
\affil{Canadian Institute for Theoretical Astrophysics, University of Toronto, 60 St. George Street, Toronto, ON M5S 3H8, Canada}
\author[0000-0002-7795-0018]{Ricardo L\'opez-Valdivia}
\affiliation{Universidad Nacional Aut\'onoma de M\'exico, Instituto de Astronom\'ia, AP 106,  Ensenada 22800, BC, M\'exico}
\author[0000-0002-0149-1302]{B\'arbara Rojas-Ayala}
\affiliation{Instituto de Alta Investigaci\'on, Universidad de Tarapac\'a, Casilla 7D, Arica, Chile }

\begin{abstract}
The fifth iteration of the Sloan Digital Sky Survey (SDSS-V) is set to obtain optical and near-infrared spectra of $\sim$5 million stars of all ages and masses throughout the Milky Way. As a part of these efforts, APOGEE \& BOSS Young Star Survey (ABYSS) will observe $\sim10^5$ stars with ages $<$30 Myr that have been selected using a set of homogeneous selection functions that make use of different tracers of youth. The ABYSS targeting strategy we describe in this paper is aimed to provide the largest spectroscopic census of young stars to-date. It consists of 8 different types of selection criteria that take the position on the HR diagram, infrared excess, variability, as well as the position in phase space in consideration. The resulting catalog of $\sim$200,000 sources (of which a half are expected to be observed) provides representative coverage of the young Galaxy, including both nearby diffuse associations as well as more distant massive complexes, reaching towards the inner Galaxy and the Galactic center.
\end{abstract}


\section{Introduction}

The Sloan Digital Sky Survey (SDSS) has obtained spectra of hundreds of thousands of stars, both within the Milky Way and beyond. Most of these are evolved stars; in particular red giants have historically been favored for the observations due to the ability to use them as tracers of stellar populations at large distances. However, SDSS has observed spectra of thousands of young stellar objects (YSOs) through its auxiliary programs \citep{roman-zuniga2023}.

During SDSS-III, young stars were targeted by the IN-SYNC program \citep{cottaar2014}. The targeting strategy  focused on known members of nearby populations visible from the northern hemisphere and containing large concentrations of young stars within the field of view of the telescope. The latter condition was required to fill a large fraction of the available spectral fibers. These included NGC 1333 \citep{foster2015}, IC 348 \citep{cottaar2015}, Orion A molecular cloud \citep{da-rio2016}, as well as NGC 2264. In total, spectra of $\sim$3,600 YSOs were taken with the APOGEE spectrograph. 
The selection strategy was based on existing catalogs of members, and thus yielded samples without significant contamination, but cannot be considered homogeneous nor complete.

SDSS-IV expanded its footprint in the targeting of YSOs and complemented previous data by observing the Orion Complex \citep{cottle2018}, the Taurus Molecular Clouds, Upper Sco, W3/W4/W5 clusters, Cygnus X, Rosette Nebula, Carina Nebula, more evolved clusters such as the Pleiades and $\alpha$ Per, and others \citep[for a complete overview of the regions observed see][]{roman-zuniga2023}. With respect to SDSS-III, a greater emphasis was given to targeting sources in a more homogeneous manner, for instance by using color cuts or photometric variability to select sources rather than relying on existing confirmation of their youth. However different selection criteria were developed for each individual region, since the observations fell under the auspices of various programs \citep{beaton2021,santana2021,roman-zuniga2023}. Spectra of $>$30,000 stars were taken across the plates covering these regions. Since many of these targets were selected prior to the release of parameters from the \textit{Gaia} mission, a significant fraction of them are evolved field stars, with the actual census of young stars within them being $<$10,000.

With the transition to SDSS-V, several changes have been implemented to the survey strategy.

\begin{itemize}
\item \textit{All sky accessibility:} the spectrographs utilized by the survey are installed on two telescopes in the northern and the southern hemisphere.
\item \textit{Fast instrument reconfiguration:} rather than using pre-drilled plates to position the fibers of the spectrographs, robotic fiber positioners are now used instead \citep{pogge2020}. This allows both a greater flexibility in targeting, and minimizing operational overheads. It is now possible to create a comprehensive sample of spectra of the stars across the Galaxy, without necessarily being limited to a particular line of sight or field of view.
\item \textit{Prioritization:} the young star program is no longer considered auxiliary; rather it is now one of the core programs of the survey.
\item \textit{Improved selection function:} a homogeneous targeting strategy across the entire sky would improve the subsequent modeling of the sample. We can no longer treat individual star forming regions independently. Regardless of the distance to a particular population, or the number of stars within it, a ``simple'' selection of all young stars in the Solar Neighborhood is needed.
\end{itemize}

However, devising an homogeneous targeting strategy is rather arduous, as, depending on their mass and age, young stars have a great degree of variety in the observational signatures that could be used to confirm their youth. As such there is no one single criterion that can uniformly select all young stars at different evolutionary stages across all masses and distances. Driven by these exigencies, we are forced to develop more sophisticated target strategies than previously implemented.  

In this paper we provide an overview of the criteria used to target the young star in SDSS-V, how it has evolved to date, and the general observation strategy. We also give a brief overview of the data collected during the first year of operations, which began in 2021.

\section{ABYSS overview}

\begin{figure*}
\epsscale{1.0}
		\gridline{\fig{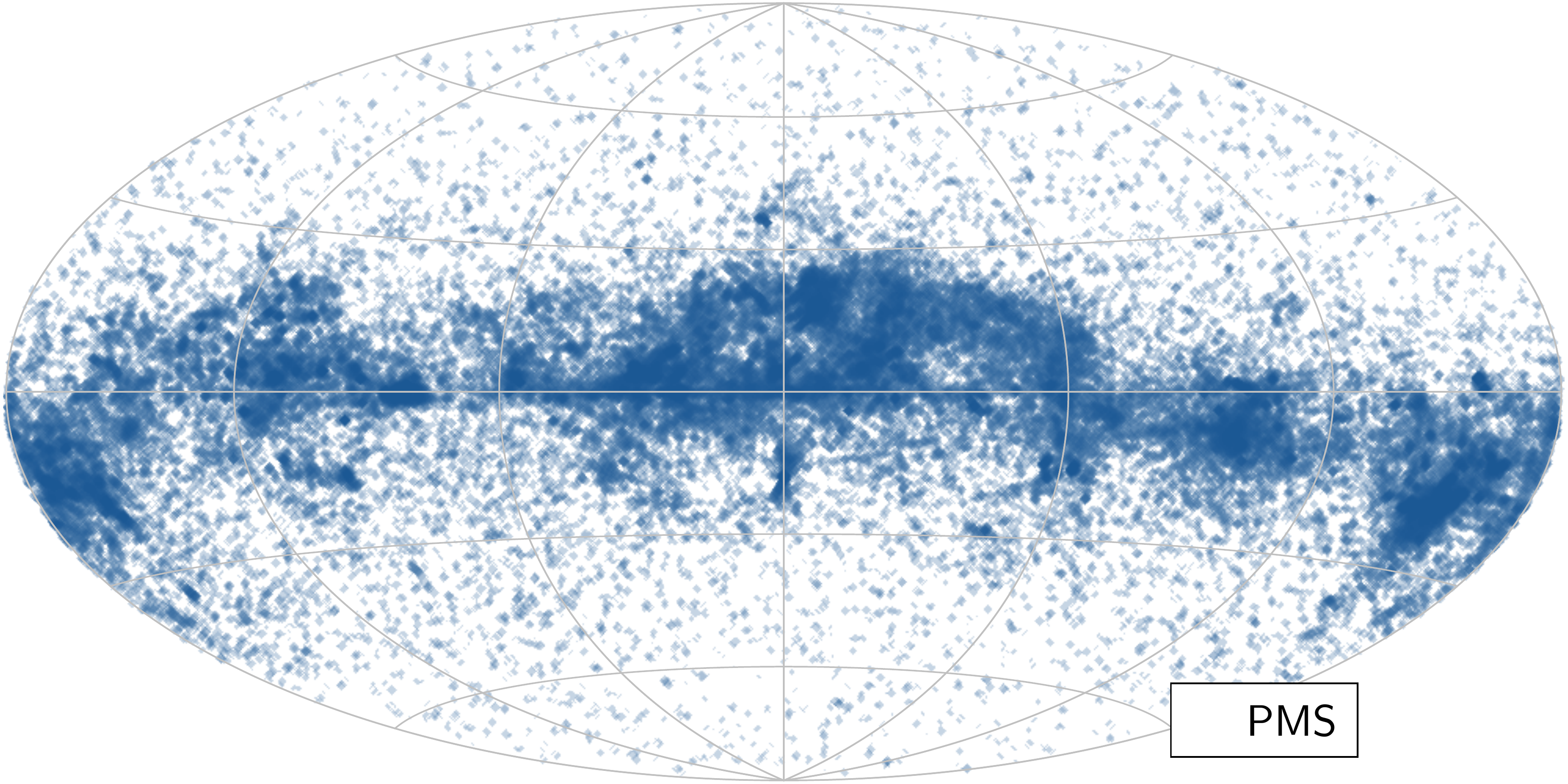}{0.5\textwidth}{}
		          \fig{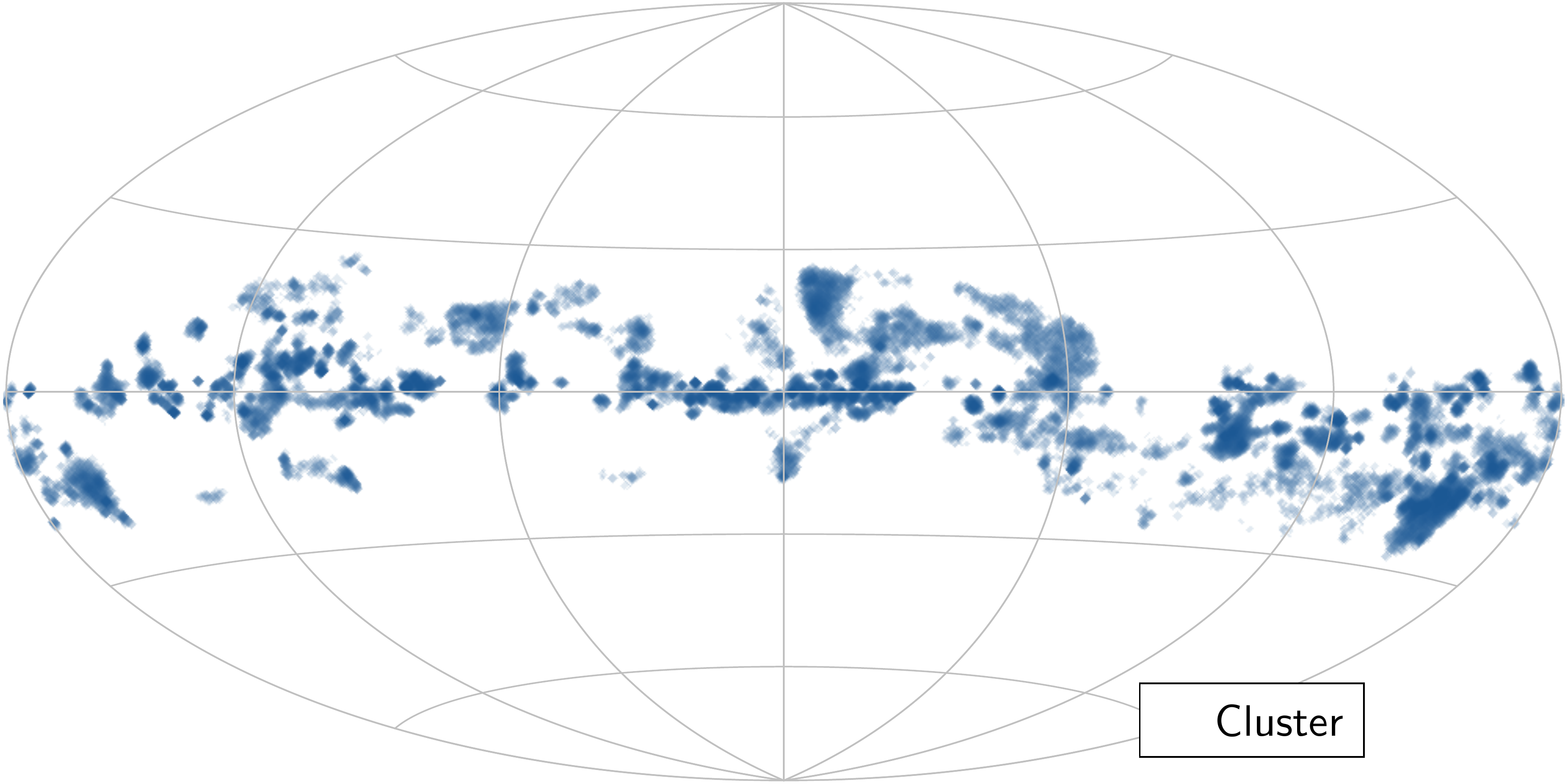}{0.5\textwidth}{}
		 }\vspace{-0.5cm}
		\gridline{\fig{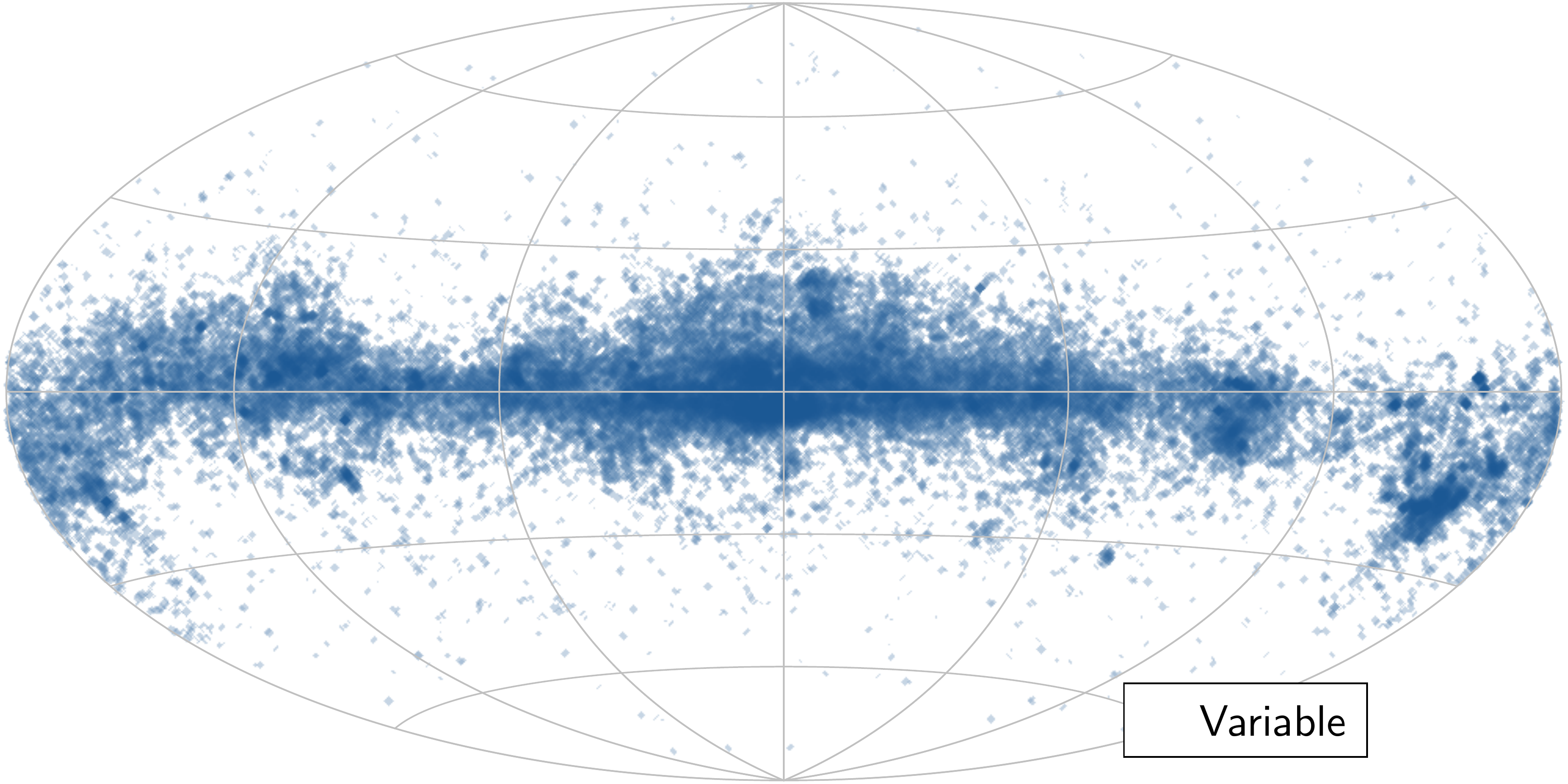}{0.5\textwidth}{}
		          \fig{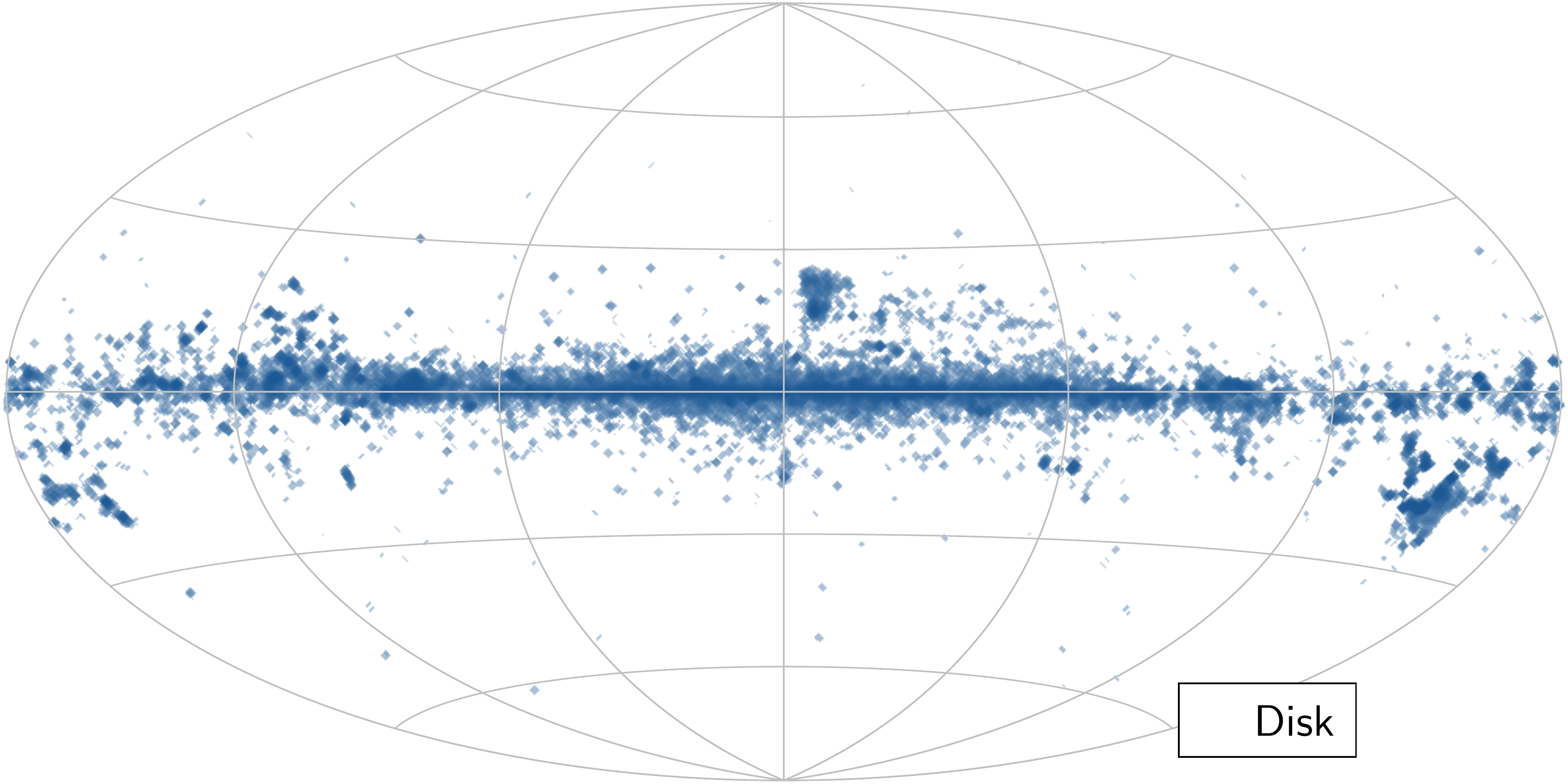}{0.5\textwidth}{}
		 }\vspace{-0.5cm}
		\gridline{\fig{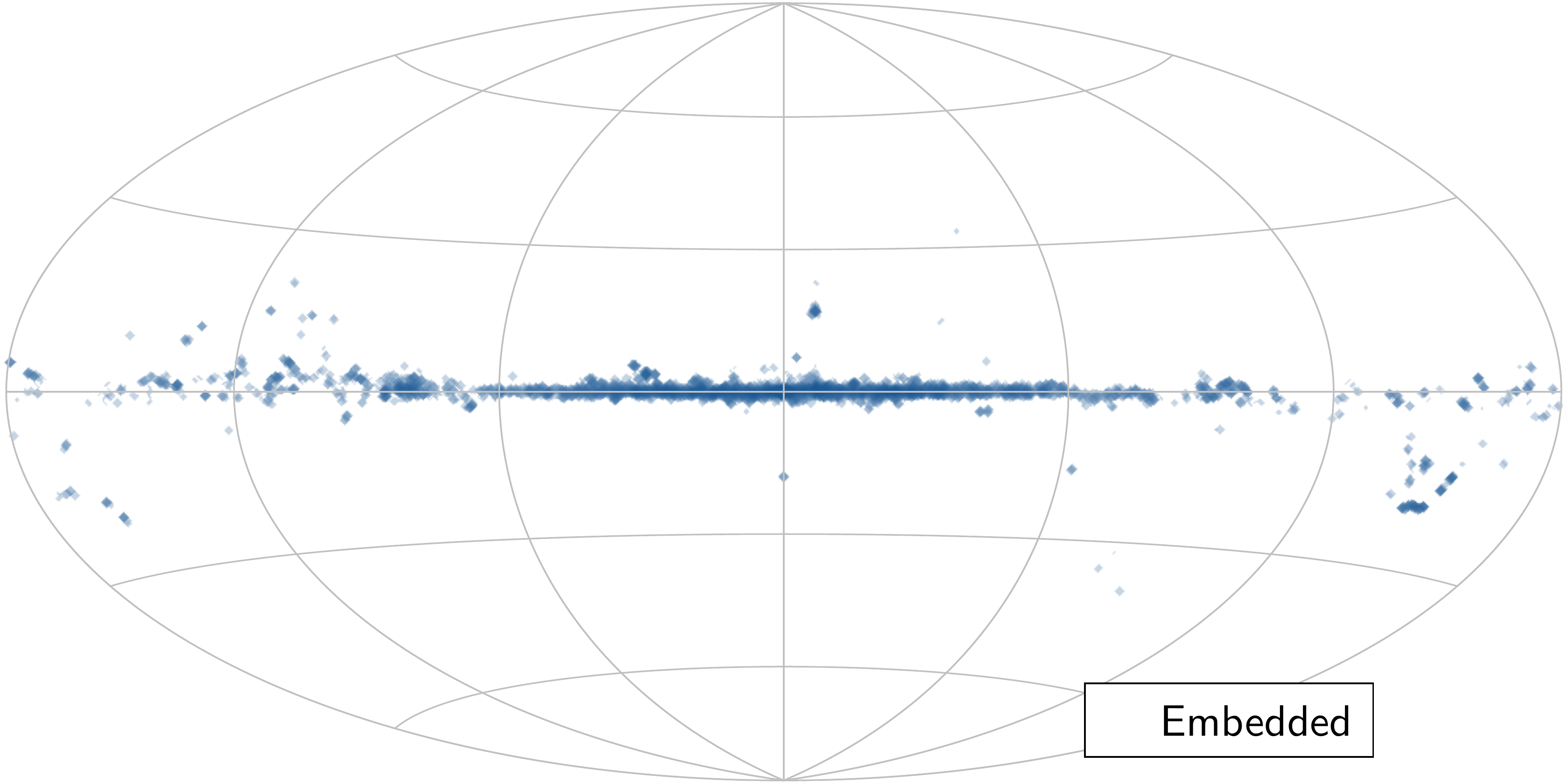}{0.5\textwidth}{}
		          \fig{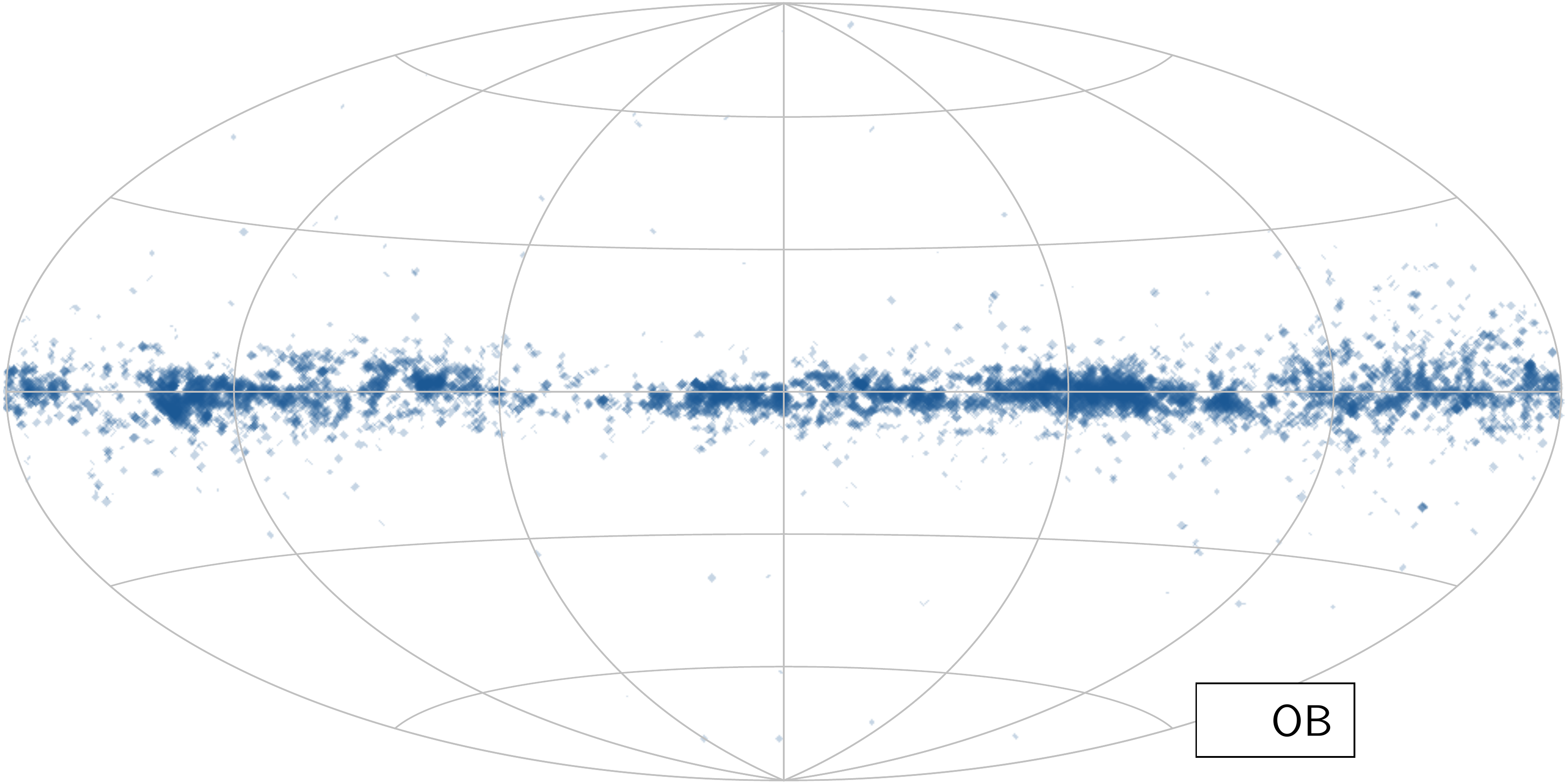}{0.5\textwidth}{}
		 }\vspace{-0.5cm}
		\gridline{\fig{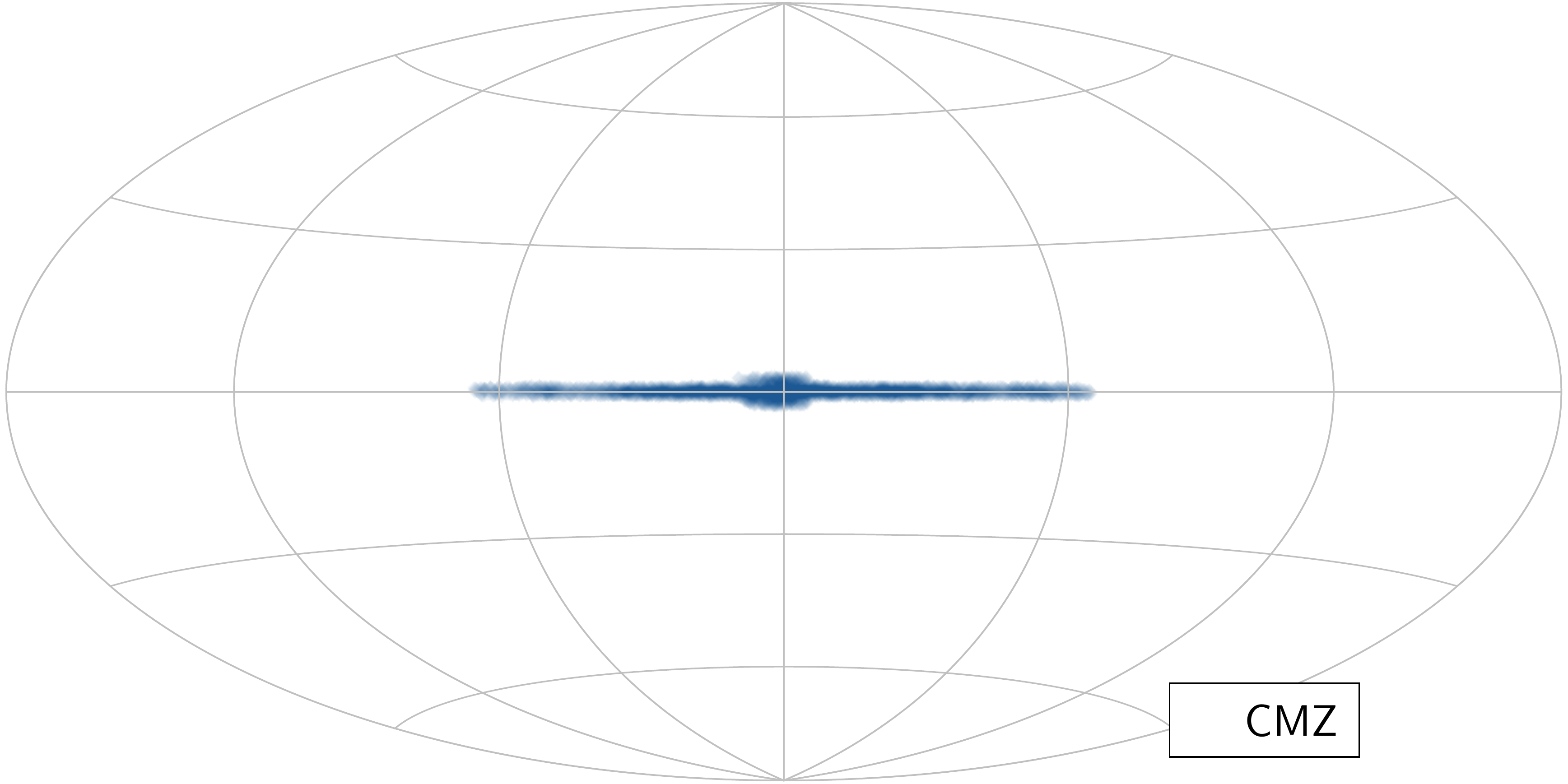}{0.5\textwidth}{}
		          \fig{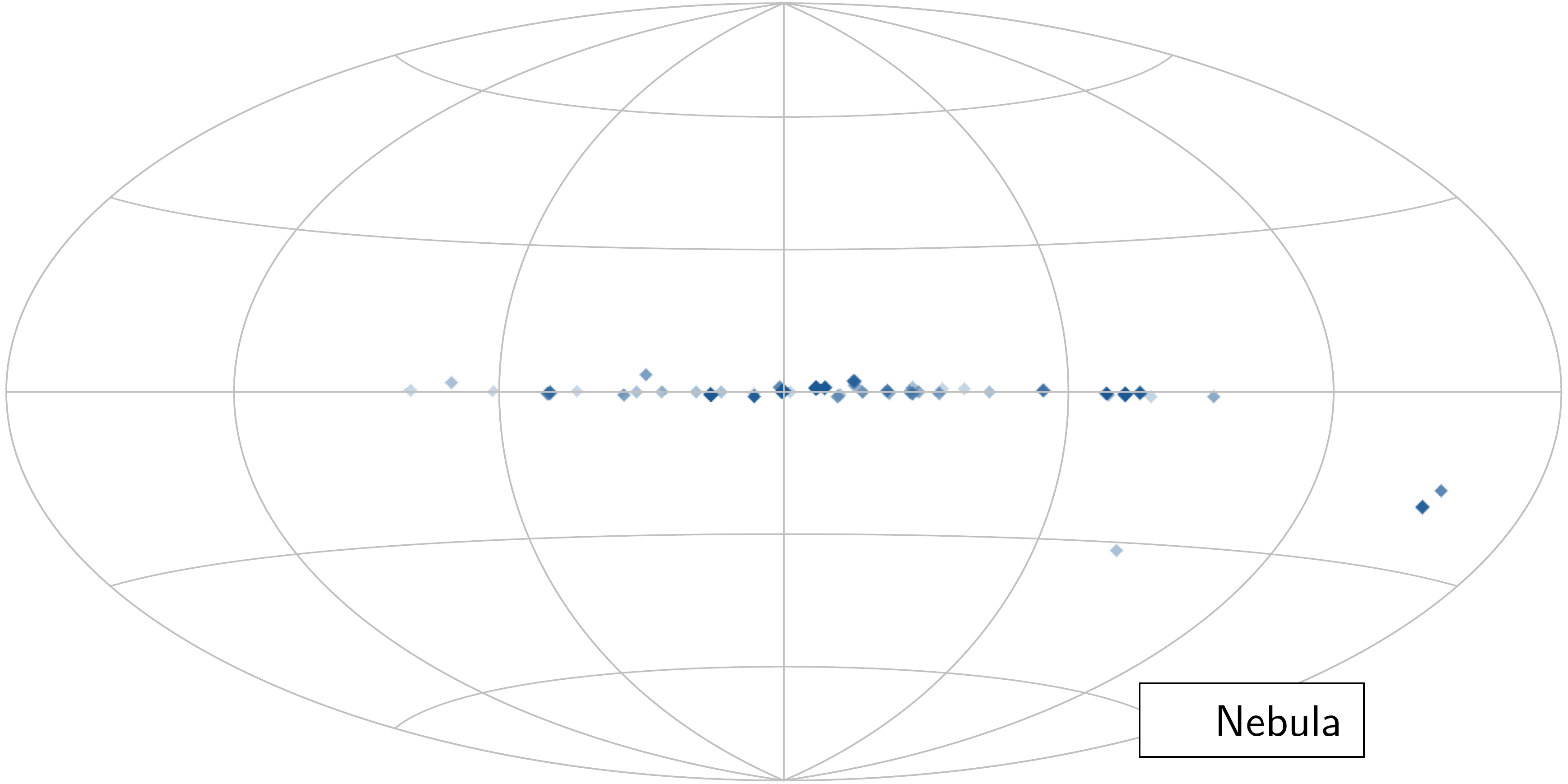}{0.5\textwidth}{}
		 }\vspace{-0.5cm}
\caption{Spatial distribution of sources in galactic coordinates for each of the cartons.
\label{fig:map}}
\end{figure*}

The APOGEE \& BOSS Young Star Survey (ABYSS) is the name of the young star program in SDSS-V. This program is set to produce optical and near-IR spectra of $\sim10^5$ young stars across the entire sky. The primary goals for these data include (but not limited to):
\begin{itemize}
    \item Characterizing the dynamical, spatial, and temporal structure of individual star forming regions, how these populations evolve over time.
    \item Tracing of the global structure traced by the young stars, from the solar neighborhood as a whole, to the Galactic scales. Characterizing the spiral arm structure, and the kinematic and dynamic properties of the young Milky Way disk.
    \item Examining the connection between young stars and gas from which they have formed.
    \item Measuring the fundamental stellar properties of young stars, their comparison to the state-of-the-art models of stellar structure and evolution. Examining the role of the environment in which a star is born on these properties.
    \item Characterizing multiplicity and orbital parameters of young stars across different populations.
\end{itemize}

In this section we present the underlying logistics behind ABYSS, including survey structure, data acquisition strategy, and target selection.

\subsection{Survey strategy}

SDSS-V utilizes two 2.5m telescopes; one is located in the northern hemisphere at the Apache Point Observatory (APO), and the second one is located in the southern hemisphere at Las Campanas Observatory (LCO) \citep{bowen1973,gunn2006,blanton2017}. Both of these telescopes have two multi-object fiber spectrographs: APOGEE \& BOSS. APOGEE covers H band, with the wavelength of 1.51--1.7 $\mu$m with R$\sim22,500$ \citep{wilson2010,majewski2017,wilson2019}. BOSS is an optical spectrograph, covering wavelength range of $\sim$3600--10400 \AA, with $R\sim$5000 \citep{smee2013}. A total of 300 APOGEE fibers and 500 BOSS fibers can be placed simultaneously in 3$^\circ$ field of view at APO, and 2$^\circ$ at LCO, with the fiber diameter of $\sim$2'' and $\sim$1.3'' respectively between these observatories.

With the new capability of SDSS to rapidly reconfigure the fiber placement due to the robotic fiber positioners \citep{sayres2021}, and the ambitious goals of a comprehensive survey obtaining spectra across entire sky \citep{kollmeier2017}, the exposure time on all fields is set at 15 minutes.

SDSS-V divides its efforts into three mappers: Milky Way Mapper (MWM), Black Hole Mapper (BHM), and Local Volume Mapper (LVM). All the programs aiming to obtain stellar spectra, including ABYSS, are a part of MWM. Each program within each mapper can define one or several ``cartons'': a subset of stars selected by a particular set of criteria that share the instrument configuration, requirements on cadence, and number of epochs.

ABYSS has defined 8 cartons of stars to be observed with APOGEE, of which 5 cartons are considered to be optically bright to also be observed with BOSS. These definitions are described in Section \ref{sec:target}. A significant fraction of targeted sources ($\sim$66\%) are observed by both instruments. This offers several advantages: precise sub-\kms\ radial velocities from high resolution APOGEE spectra, versus various lines that can clearly inform on stellar youth (such as Li I and H$\alpha$) that can be accessed with BOSS. The two instruments have different faint limits set by the program: $H<13$ mag for APOGEE and $G_{RP}<15.5$ mag for BOSS. The faint limit is set at the typical magnitude that would reach signal-to-noise ratio of 30 in a coadded spectrum of three 15 minute exposures, which was deemed sufficiently high to extract the fundamental stellar parameters from the spectra.

These limits apply for all of the defined cartons. As some of the YSOs are too faint to be detected in the optical regime due to extinction, some of the cartons can only be observed with APOGEE. Individual stars can also meet the faint limit in optical, or in infrared, but not both, as such, $\sim$20\% of the stars for either spectrograph are unique, with the remaining $\sim$80\% being targeted by both instruments.

In total for ABYSS targets, a ``complete'' set of observations requires 3 APOGEE and 3\footnote{For fainter optical targets, 4 BOSS epochs are requested for sources with 14.76$<G_{RP}<$15.075 mag, 5 epochs for 15.075$<G_{RP}<$15.29 mag, and 6 epochs for 15.29$<G_{RP}<$15.5 mag.} BOSS epochs for a given object,  brightness limits permitting. Such a number of exposures is needed to confirm multiplicity or variability within a spectrum. No firm constraints on the cadence of observations have been imposed. 

\subsection{Carton definitions}\label{sec:target}

In this subsection we present 8 independent definitions for cartons that were used to target sources for observations as part of the ABYSS program (Figure \ref{fig:map}). These cartons rely on various criteria, such as infrared excess, position on the HR diagram, photometric variability, and membership of moving groups/clusters/etc.

The initial set of observations conducted during the first 6 months of operations, still using plug plates (see Section \ref{sec:firstyear}) relied on the V0 version of the targeting. With the instrument upgraded to robotic positioners, the targeting criteria were updated to V0.5. Transition to V1 will occur in 2023. Data release 18 makes available V0.5 version of the targeting, which is the focus of this paper. However, for the sake of the historical record, the full evolution of the selection is described.

\subsubsection{Disk}

\begin{figure*}[!ht]
\epsscale{1.0}
		\gridline{\fig{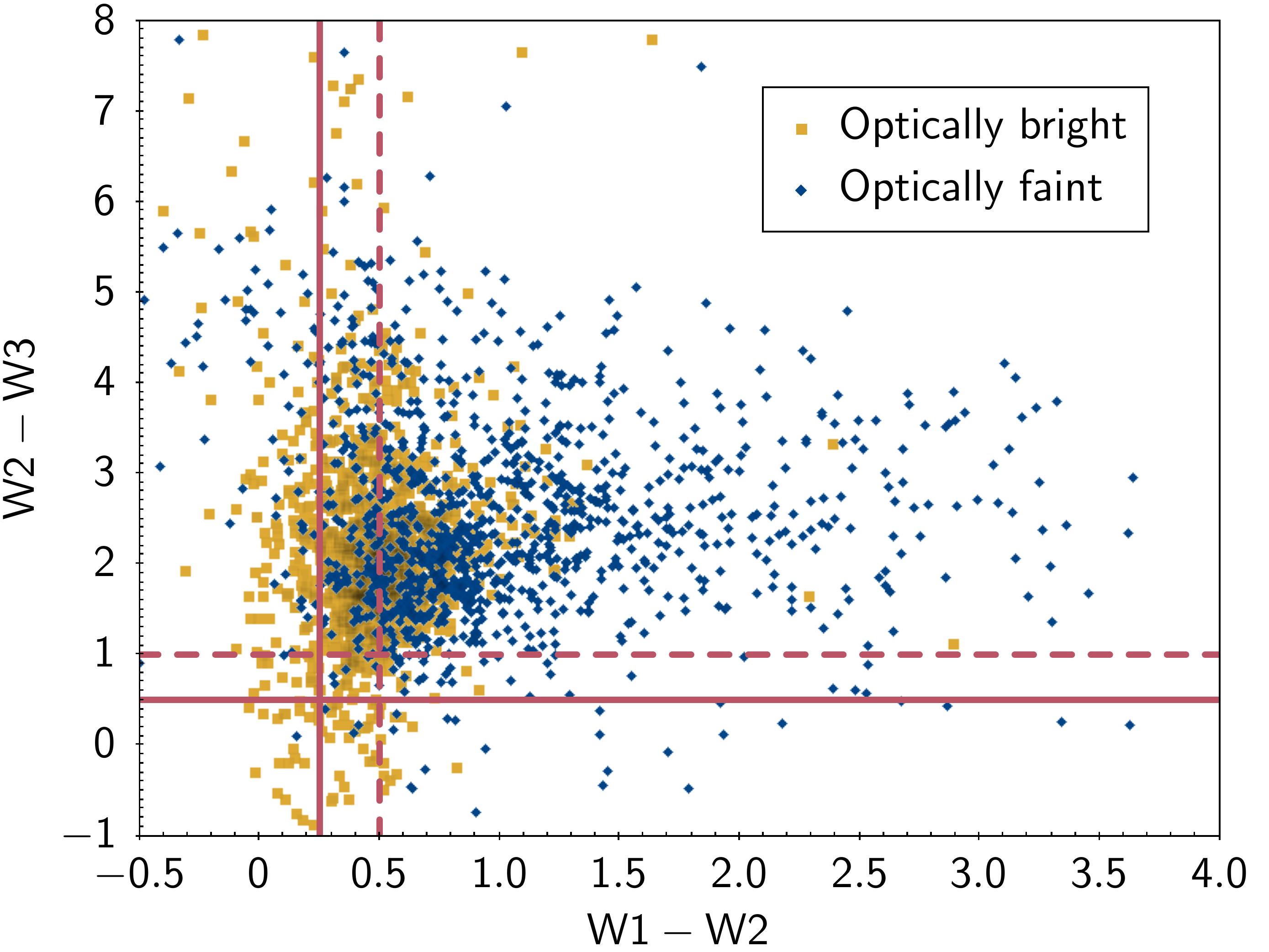}{0.5\textwidth}{}
		          \fig{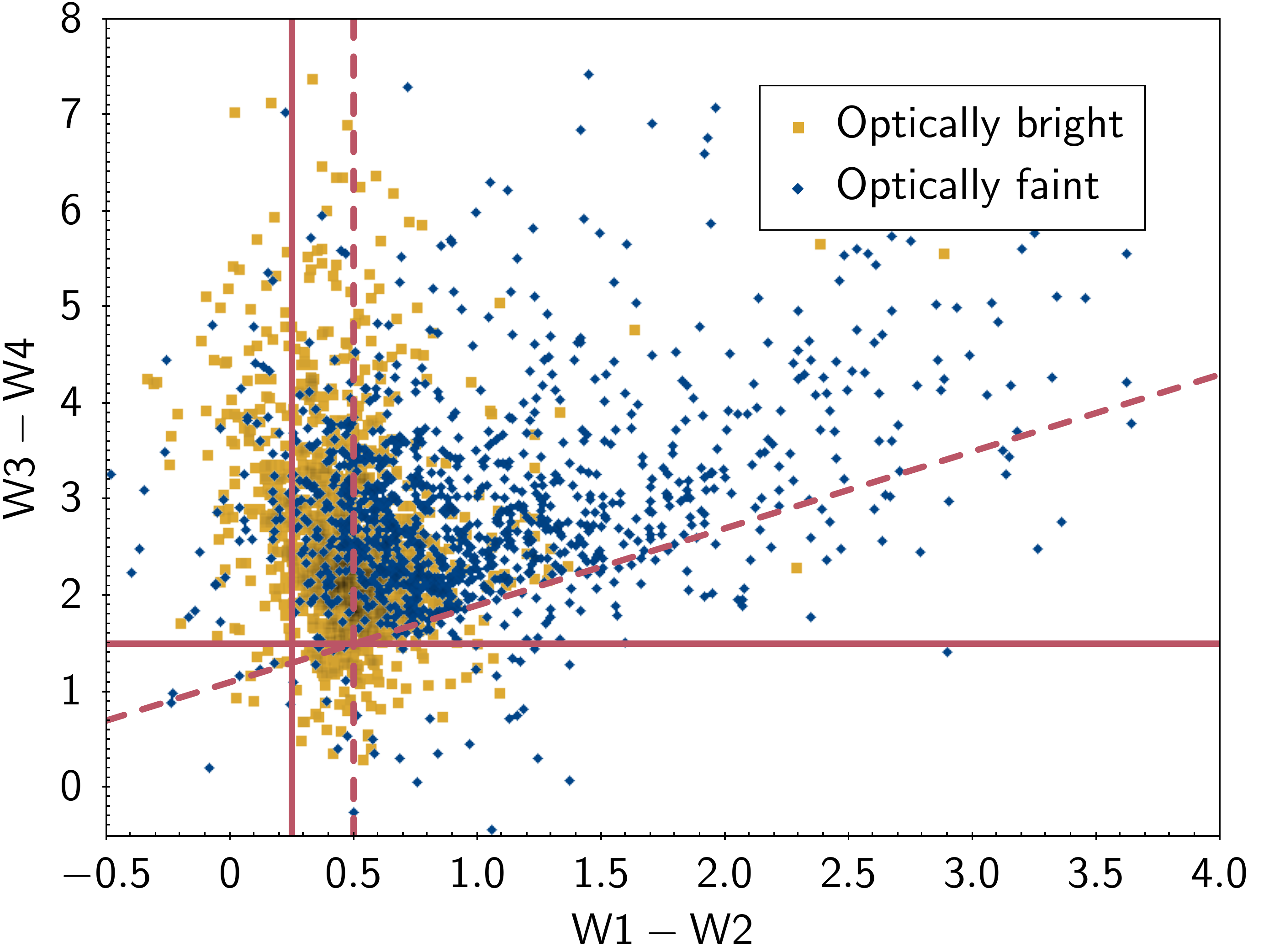}{0.5\textwidth}{}
		 }\vspace{-0.5cm}\gridline{
		          \fig{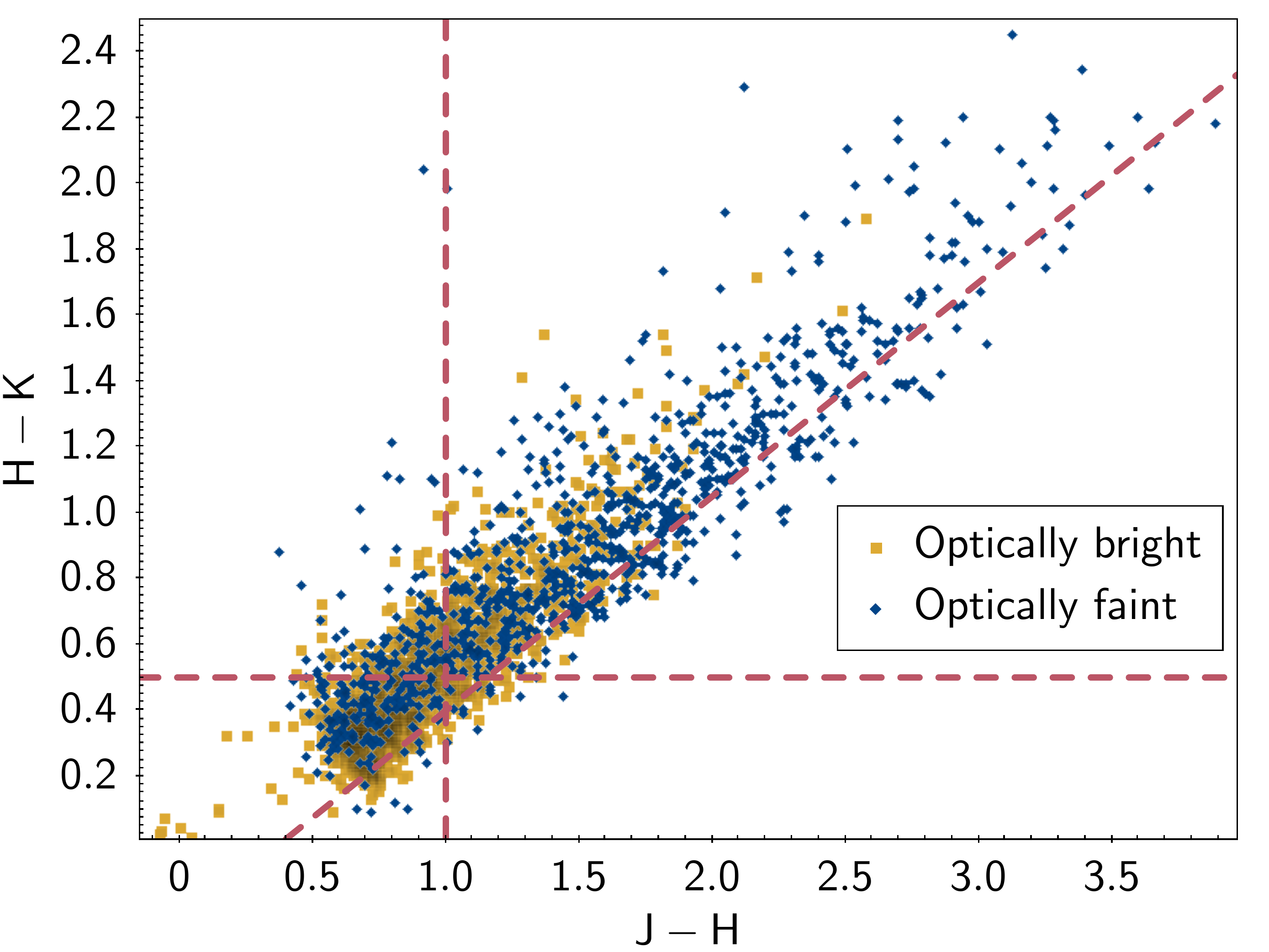}{0.5\textwidth}{}
		          \fig{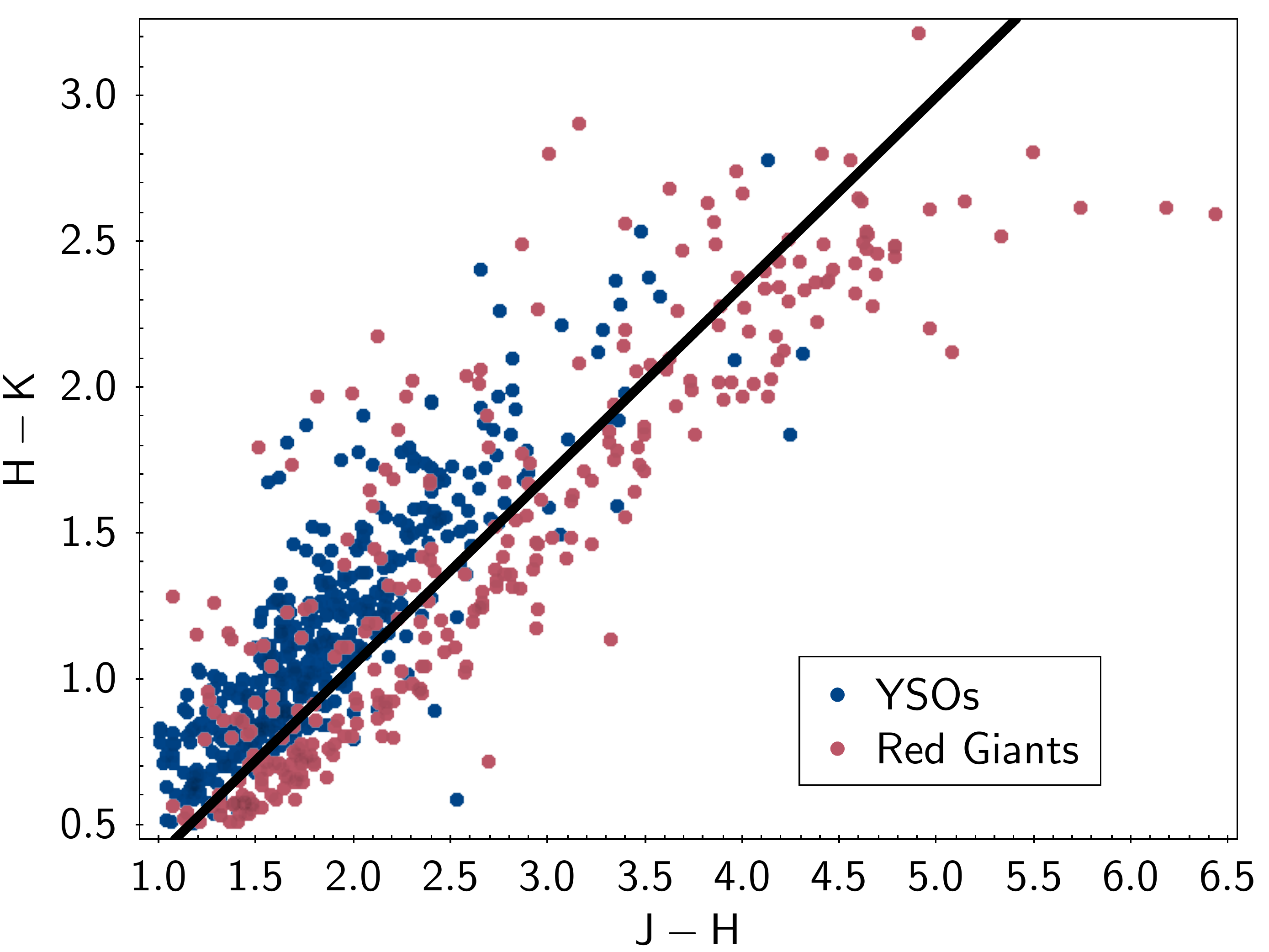}{0.5\textwidth}{}
        }\vspace{-0.5cm}
\caption{Criteria for the Disk and Embedded cartons, showing color-color diagrams of disk-bearing dusty YSOs from \citet{megeath2012} in Orion A \& B molecular clouds, used as a reference to constrain the targeting selection. Sources in yellow are those with $G<18.5$, $\pi>0.3$ mas, used as a template to map the Disk carton; the relevant color cuts to select this carton are shown as red solid lines. Sources in blue are optically faint, representative of the Embedded carton; the color cuts are shown in red dashed lines. Bottom right panel shows the data for sources obtained as part of ABYSS observations with APOGEE, separating the sources into likely YSOs and the contaminating red giants through \logg$>$3.2 cut. The black line shows the color cut introduced for the Embedded carton in the V1 targeting, as prior to this it had significant contamination.
\label{fig:dusty}}
\end{figure*}

Historically, the YSOs that have been easiest to identify are those that have large infrared excess due to the presence of a protoplanetary disk, particularly in the mid-IR regime. In the last few decades, telescopes such as Spitzer and WISE have particularly expanded the census of dusty YSOs. In particular, WISE, due to being an all-sky survey, is particularly informative for targeting. Several studies have used WISE to search for YSOs \citep[e.g.][]{koenig2014, kang2017, marton2016}, but they either focused only on a specific star-forming region, or they had a large degree of contamination across the entire sky. 

As at this stage in the survey, the goal is to create a census of sources that should be targeted for follow-up observations (rather than explicit classification of YSOs into evolutionary stages), thus we use simple color cuts in WISE photometry for this carton. To minimize a selection of very distant, highly extincted field stars (which are the main source of contamination), we impose a parallax cut as well - this implicitly requires all of the identified sources to be bright enough in the optical regime to be detected and have reliable astrometry with Gaia. We select sources satisfying:

\begin{itemize}
\item $W1-W2>0.25$ mag
\item $W2-W3>0.5$ mag
\item $W3-W4>1.5$ mag
\item $\pi>0.3$ mas.
\end{itemize}
\noindent These cuts have been evaluated against known dusty YSOs in the Orion Molecular Clouds \citep{megeath2012}, and they are shown in Figure \ref{fig:dusty}.

ALLWISE photometry was used for the selection. Although there have been recent re-reductions, such as unWISE or neoWISE, their improvements are primarily in W1 and W2 band photometry, W3 and W4 mostly remain as is. Longer wavelength bands lack the sensitivity of shorter wavelength bands, furthermore, they have not been observed for as long due to WISE running out of cryogenic coolant needed to suppress telescope emission at these wavelengths. Nonetheless, W3 and W4 bands are critical for reliably identifying dusty disk-bearing stars. As such, by requiring these bands and using merged photometry, any improvements in W1 or W2 have negligible effect on the selection.

In the V0 version of the targeting, these cuts formed the basis of YSO\_S1 carton. In version V0.5, the carton was renamed and split into YSO\_Disk\_APOGEE and YSO\_Disk\_BOSS, containing 28,832 and 37,478 stars respectively. In version V1, Gaia DR2 astrometry was upgraded to EDR3.

\subsubsection{Embedded}

Some disk-bearing sources are too faint to have reliable Gaia parallaxes. This is usually the case for Class I protostars that are still embedded in their natal envelopes, Class II YSOs that have edge-on disks, or for the sources that are more distant and thus have more extinction along the line of sight. Without a distance estimate, it can be difficult to separate bona-fide YSOs from distant and heavily extincted red giants. As such, more stringent color cuts, not just on ALLWISE photometry but also 2MASS, are required.

\begin{itemize}
\item $G>18.5$ mag, or undetected
\item $J-H>$1 mag, $H-K>$0.5 mag
\item $W1-W2>0.5$ mag, $W2-W3>1$ mag, $W3-W4>1.5$ mag
\item $W3-W4>0.8\times(W1-W2)+1.1$ mag
\end{itemize}
\noindent These cuts are shown in Figure \ref{fig:dusty}.

In V0 version of targeting, the carton was referred to as YSO\_S2; in V0.5 it was renamed as YSO\_Embedded\_APOGEE, containing 11,086 stars. Following the first year of operations, the carton was re-examined; approximately half of the observed sources were red giants, as shown in Fig. \ref{fig:dusty} (bottom right panel). Therefore in V1 we added an additional cut
\begin{itemize}
\item $H-K>0.65\times(J-H)-0.25$ mag
\end{itemize}
to minimize the contamination by evolved stars and further restrict the selection to the parameter space that is most commonly inhabited by spectroscopically confirmed YSOs, limiting the sample to 5,455 stars.

\subsubsection{Nebula}

\begin{figure}
\epsscale{1.1}
\plotone{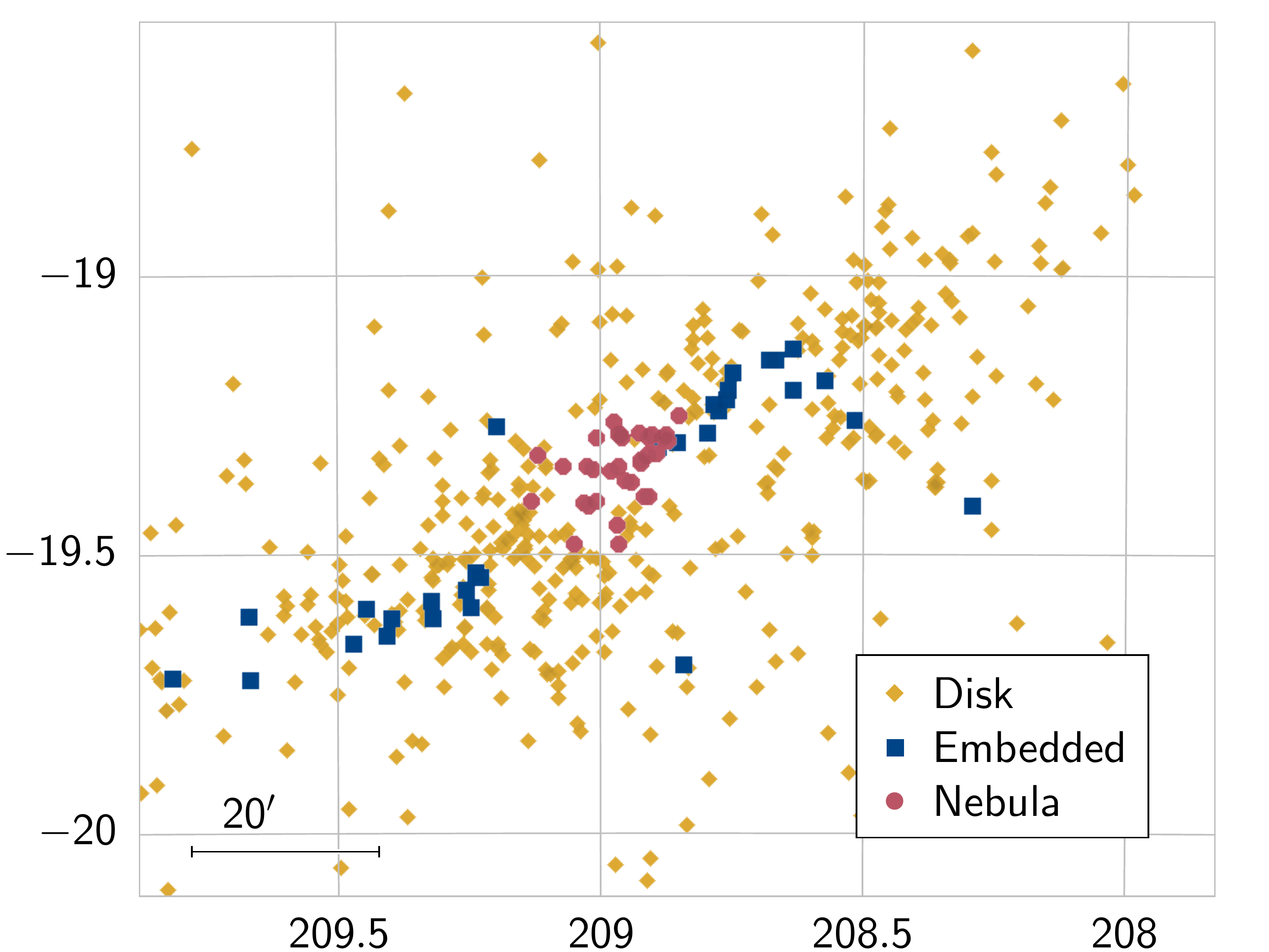}
\caption{Spatial distribution (in Galactic coordinates) of the selected sources in Disk, Embedded, and Nebula cartons toward the Orion Nebula Cluster. Note that the sources in the Nebula carton fill in a gap in the other two cartons.
\label{fig:nebula}}
\end{figure}
The selection criteria of disk-bearing and embedded sources rely on WISE bands $W3$ and $W4$. $W3$ and $W4$ however become less effective in regions of high nebulosity, as they saturate, producing gaps in the coverage. To fill them, we used shorter wavelength data, using a set of criteria that is tuned to autonomously find such nebulous regions:
\begin{itemize}
\item If W4 is not reported, $W2-W3>4$ mag
\item If W3 and W4 are not reported, $J-H>1.1$ mag
\end{itemize}

This preferentially selects sources found in gaps of the previous cartons in discrete regions on the sky, such as, e.g., in the center of the Orion Nebula (Figure \ref{fig:nebula}).

Additionally, some of the sources that are selected by these cuts are found off of the Galactic plane and/or away from known star forming regions. This creates an excess of targets in a narrow line following the scanning law of WISE telescope, as such, these sources appear to be suspect. Thus, we also required $b<5^\circ$ and a combination of $b>-5^\circ$ or $l>180^\circ$, to exclude this contamination.

This carton was introduced in V0 as YSO\_S2.5. In V0.5 it was renamed as YSO\_Nebula\_APOGEE, containing 1,112 stars.

\subsubsection{CMZ}

The inner Galaxy, including the central molecular zone (CMZ), has been surveyed with Spitzer as a part of GLIMPSE and MIPSGAL programs \citep{churchwell2009,carey2009,gutermuth2015}. These data offer a substantial improvement on sensitivity and resolution in comparison to WISE, therefore they are advantageous in targeting stars outside of the Solar Neighborhood.

Using the properties of massive YSOs towards the Galactic center identified by \citet{an2011}, we select a sample of candidates with

\begin{itemize}
\item $[8.0]-[24]>2.5$ mag, i.e., very red sources, using photometry from \citet{gutermuth2015}.
\item $\pi<0.2$ mas or not detected/measured, to ensure the sources are distant.
\end{itemize}

In V0 of targeting, YSO\_CMZ carton also imposed a spatial limit of $358<l<2^\circ$ and $-1<b<1^\circ$ to focus solely on the central molecular zone. In V0.5, the carton was renamed to YSO\_CMZ\_APOGEE and removed the spatial restriction, allowing the sources from the entire MIPSGAL footprint. This enables to identify YSO candidates across the inner Galaxy, containing 13,170 stars.

\subsubsection{Variable}\label{sec:variable}

\begin{figure*}
\epsscale{1.0}
		\gridline{\fig{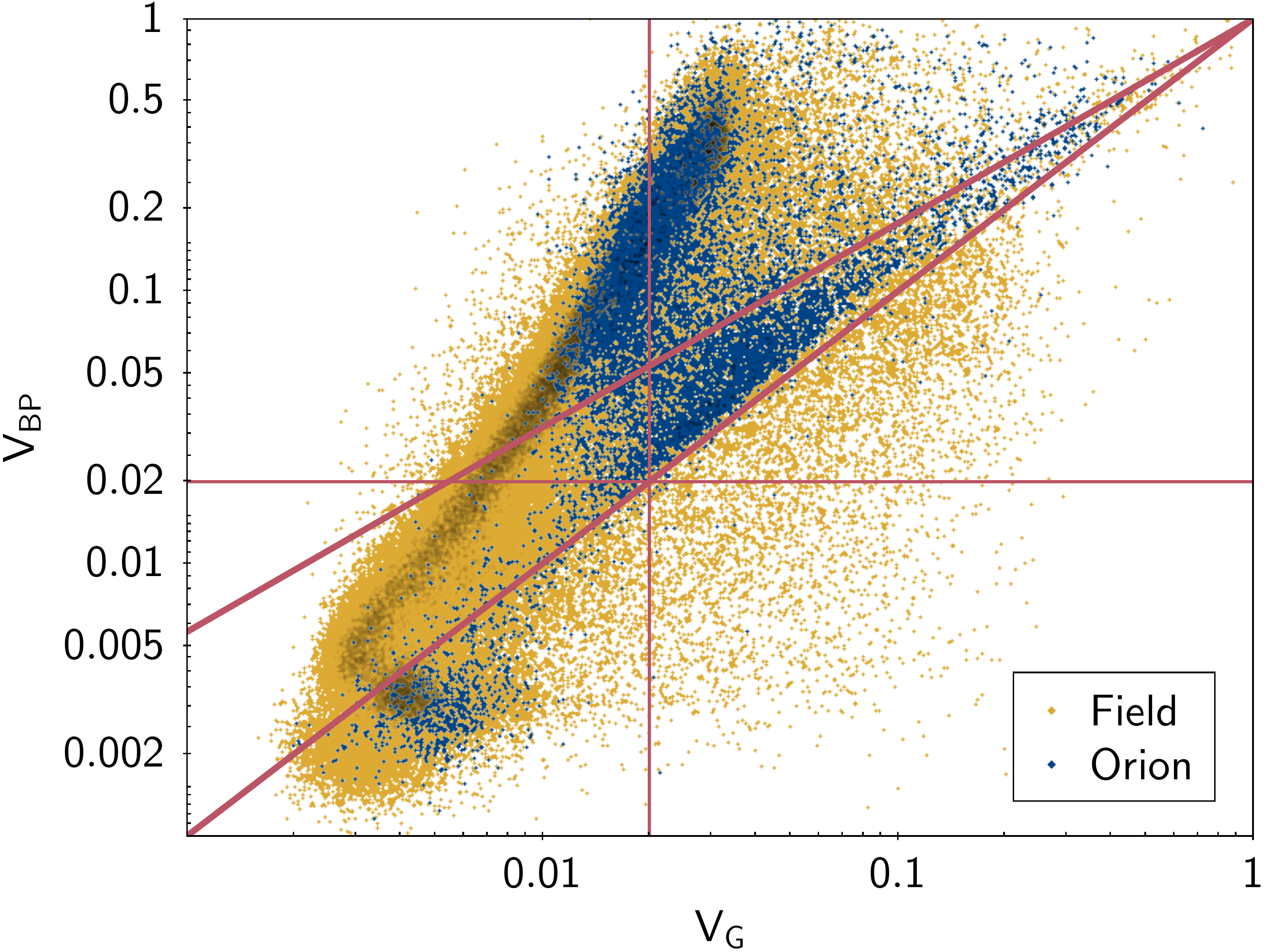}{0.5\textwidth}{}
		          \fig{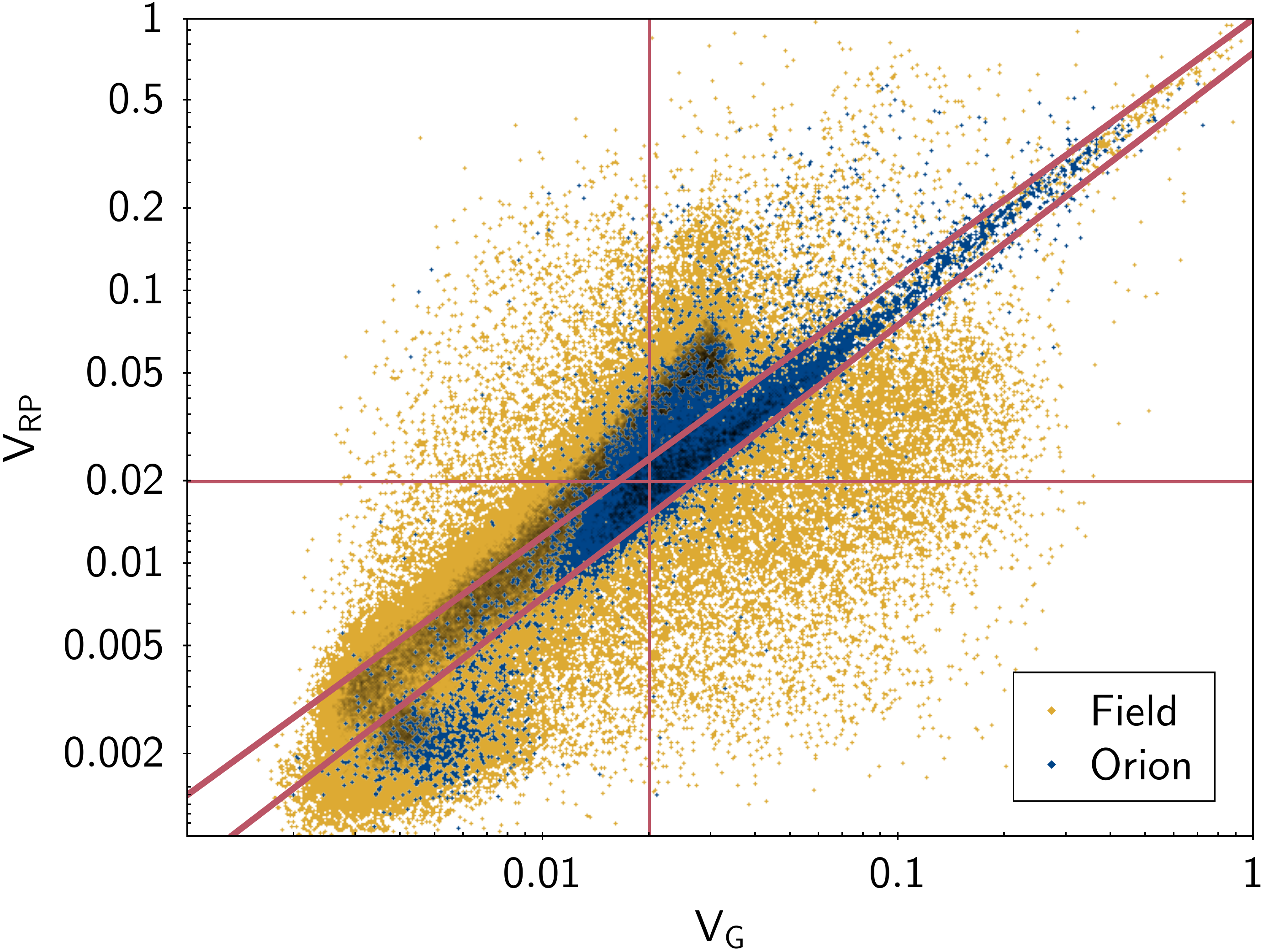}{0.5\textwidth}{}
		}\vspace{-0.5cm}\gridline{
		          \fig{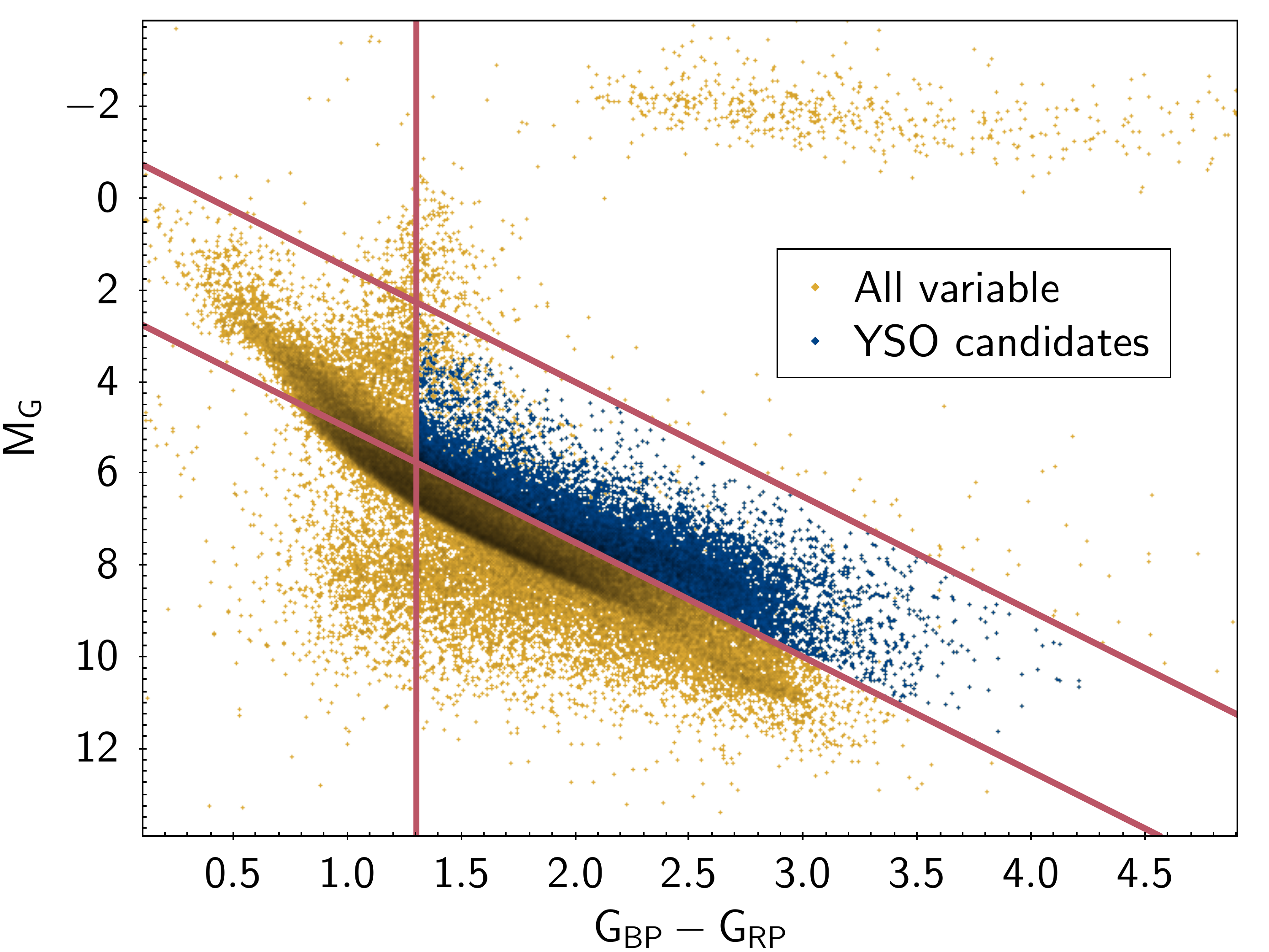}{0.5\textwidth}{}
        }\vspace{-0.5cm}
\caption{Criteria for the selection of stars in the Variable carton. Top panels show variability distribution in the sample of stars toward the Orion Complex, with known members highlighted in blue, and field stars shown in yellow. The red lines show the cuts to the sample based on the correlation of variability on the order of unity described in the text. Bottom panel shows the full sample of stars within 500 pc that meets the minimum variability and variability correlation cuts (See Section \ref{sec:variable}), showing the sources that have been selected as YSO candidates in blue, and the other variable stars in yellow.
\label{fig:var}}
\end{figure*}

On average, YSOs tend to be more variable than main sequence stars \citep[e.g.,][]{kounkel2022a}. Part of the reason for this variability is the presence of stronger magnetic fields that leads to more prominent star spots with a larger filling factor. Also the presence of protoplanetary disks that would occult the photosphere, and accretion events can lead to an increase in brightness.

To estimate variability $V_x$, we use multi-epoch photometry by Gaia. Following \citet{belokurov2017}, we define the photometric variability in a given filter $x$ as:
\begin{equation*}
V_x=\sqrt{\mathrm{phot\_x\_n\_obs}}/\mathrm{phot\_x\_mean\_flux\_over\_error},
\end{equation*}
where $x$ corresponds to the \textit{Gaia} $G$, $G_{BP}$, or $G_{RP}$ bands, $\mathrm{phot\_x\_n\_obs}$ is the number of observations which contributed to the photometry in a given band, $\mathrm{phot\_x\_mean\_flux\_over\_error}$ is the mean flux in a given band divided by its error. For strongly variable sources, the photometric uncertainty is comparable to the amplitude of variability.

Using the list of members of the Orion Complex from \citet{kounkel2020} as a representative sample of young stars, we develop a set of criteria based on variability that allows to most cleanly preserve a large fraction of members while rejecting field stars within the volume of space surrounding Orion. This set of criteria was later applied to the entire sky and further modified to preserve the morphology of nearby star forming regions and minimize contamination, resulting in the following:

\begin{itemize}
\item $V_G>0.02$, $V_{BP}>0.02$, $V_{RP}>0.02$. The reference YSOs tend to be more variable than the field stars. 
\item $V_G^{0.75}<V_{BP}<V_G$, $0.75V_G<V_{RP}<V_G^{0.95}$. Correlation of variability in different bandpasses on the order of unity appears to be a strong indicator of YSOs. On the other hand, while YSOs with different correlation in variability do exist, it is difficult to reliably separate them from the field stars, as such many young variable stars may be excluded from the selection. The slopes of these power laws were determined from examining Figure \ref{fig:var}.
\item $G_{BP}-G_{RP}>1.3$. Hot stars do not have convective atmospheres, and thus they generally don't have spotted photospheres. While this specific cut is redder than the convective limit, it also minimizes the extreme contamination from the red giants. Variability among hotter stars may often be an indicator of other processes, such as, e.g., eclipsing binaries, which is not an indicator of youth.
\item $M_{BP}>5\log_{10}{V_{BP}}+11$ - to preferentially exclude the evolved subgiants, despite some overlap in the parameter space with bona-fide YSOs.
\item $2.5(G_{BP}-G_{RP})-1<M_G<2.5(G_{BP}-G_{RP})+2.5$. Previous criteria preferentially select YSOs, but it still includes some strongly variable main sequence stars. This selection confines the parameter space on the HR diagram to minimize contamination.
\item $\pi>0.3$ mas, $G<18.5$, $H<13$, to limit the selection to the sources to brighter stars with reliable parallaxes. Note that $H$ band cut is applied both to APOGEE and to BOSS sample, as sources with $H>13$ \& $G_{RP}<15.5$ preferentially trace out more distant stars that seem to be more strongly contaminated that the sources within 1 kpc. As such, BOSS variable sample is a subset of the APOGEE variable sample, with the faint limit in place.
\end{itemize}
\noindent These cuts are shown in Figure \ref{fig:var}.

The carton based on this selection has been introduced in V0 as YSO\_S3. In V0.5 it has been renamed and split into YSO\_Variable\_APOGEE and YSO\_Variable\_BOSS, containing 52,691 and 47,758 stars respectively. In V1, the selection has been upgraded from Gaia DR2 data onto Gaia EDR3.

We note that Gaia DR3 has produced a catalog of young star candidates based on their variability \citep{marton2022}. The selection presented here was originally derived prior to the availability of these data. Out of 79,375 sources presented in Gaia DR3, our selection has 3,963 stars in common with this carton, and 16,474 stars across all of the cartons. Of the remaining 20,225 candidates in Gaia DR3 that would meet our faint limit, as much as a half appear to be contamination from highly reddened distant main sequence and red giant stars, but, in future versions of the targeting definition, with some refinement it may be possible to take advantage of this catalog. On the other hand, our current selection does not extend to as faint magnitudes (and, indeed, applying the same criteria to fainter stars does appear to significantly increase contamination), but it does appear to have greater sensitivity to the populations with ages of up to a few 10s of Myr.

\subsubsection{Cluster}
Most of the selection criteria devised in this work preferentially target low-mass YSOs, still in the pre-main sequence (PMS) phase of stellar evolution. Young late B, A, \& F stars reach the main sequence quickly and become difficult to separate from field stars using conventional photometric selection criteria. To identify them, it is however possible to take advantage of the fact that young stars generally form in large associations, typically with hundreds or thousands other members. Young populations tend to be dynamically cold, with velocity dispersion of less than a few \kms. Thus, it is possible to find young moving groups by performing clustering analysis in position and velocity phase space. As a selection of likely members using this method does not have a dependence on the spectral type, it makes possible to include young B, A, and F stars alongside later type stars.

\citet{kounkel2020} have applied hierarchical clustering on Gaia DR2 data within 3 kpc of the Sun. The initial data selection consisted of $\pi>0.2$ mas, $-30<b<30^\circ$, $v_{\alpha,\delta}^{lsr}<60$ \kms, as well as additional cuts based on astrometric and photometric quality. The clustering was performed with HDBSCAN \citep{hdbscan1} in several slices in distance and then stitched together. In total more than 8,000 moving groups were identified consisting of $\sim$1 million stars. The ages of the identified moving groups were estimated through an isochrone fitting using a neural net Auriga \citep{kounkel2020}.

We selected all sources in the moving groups with an age $\log t \mathrm{(Myr)}<7.5$. The resulting subset forms the basis of YSO\_Cluster carton, introduced in V0 version of targeting, and split into YSO\_Cluster\_APOGEE and YSO\_Cluster\_BOSS in V0.5, containing 45,461 and 59,065 stars respectively.

Older populations, with $\log t \mathrm{(Myr)}>7.5$ are being considered by the survey as a part of an Open Fiber Program, but only as targets of opportunity, with a single epoch obtained with either BOSS or APOGEE, and are not included among the core programs of the survey.

\begin{figure}
\epsscale{1.2}
\plotone{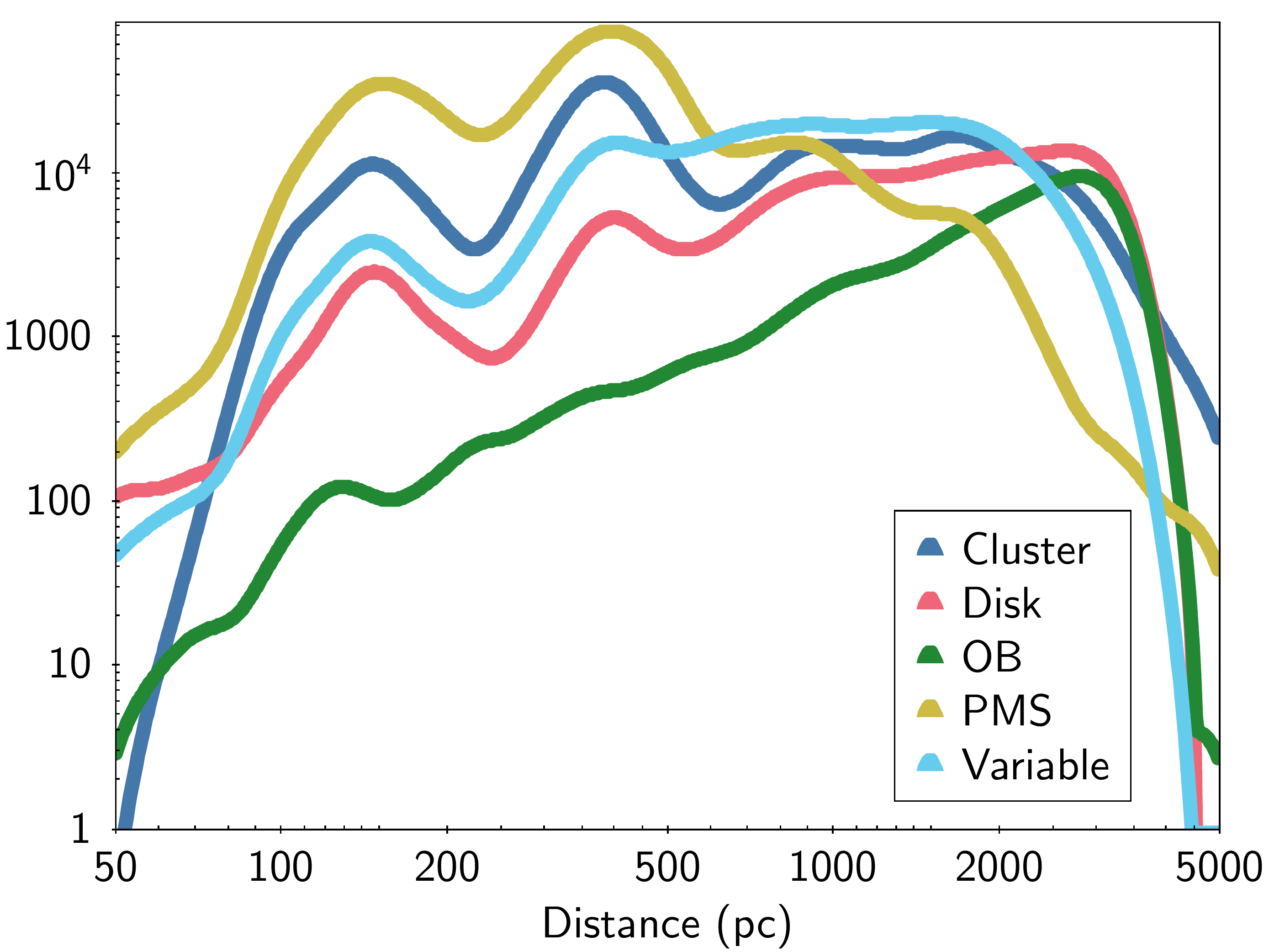}
\caption{Distribution of distances toward the stars in the optically bright cartons.
\label{fig:dist}}
\end{figure}

\begin{figure}
\epsscale{1.0}
\plotone{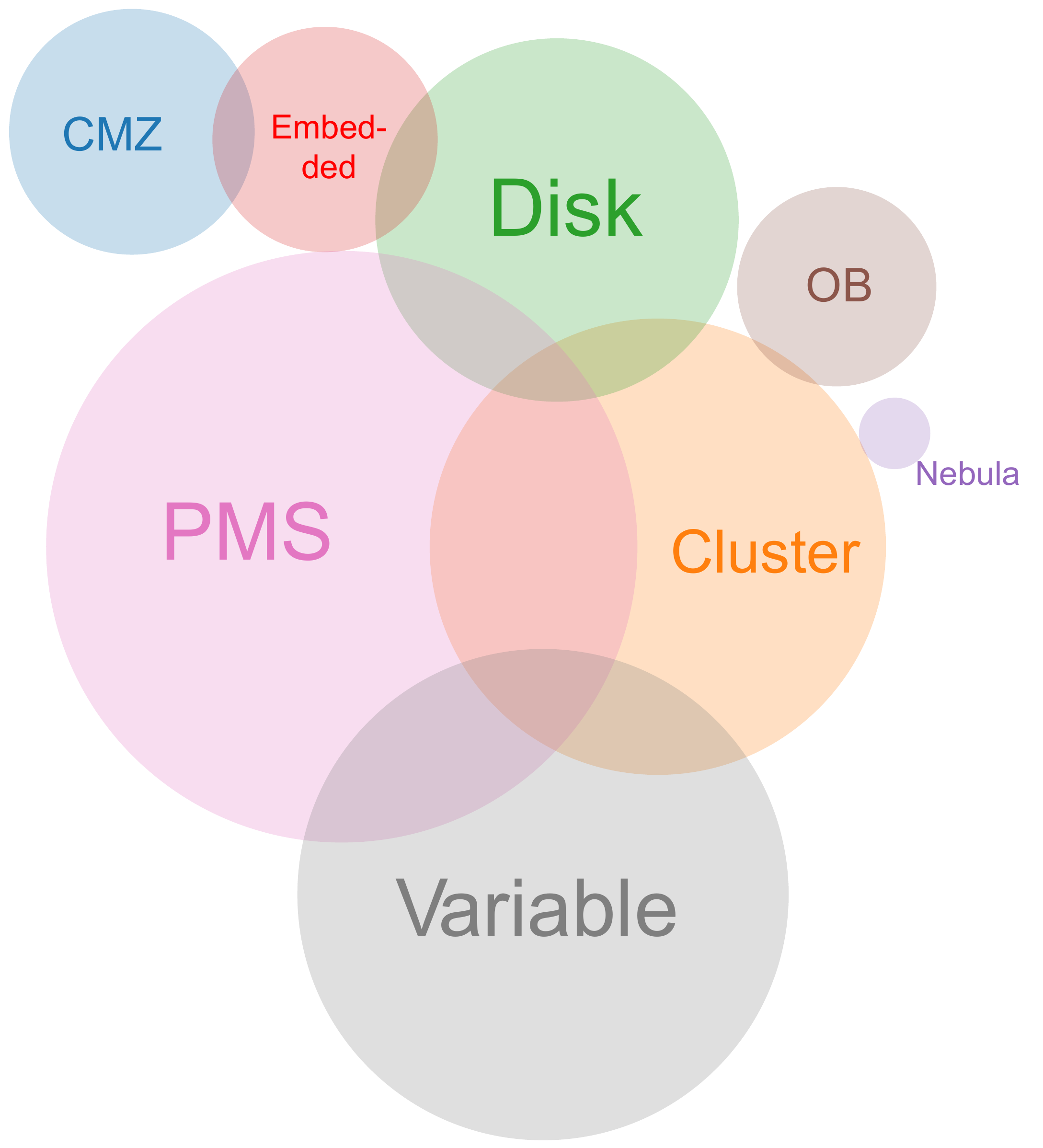}
\caption{A Venn diagram showing relative sizes of all the various cartons in the ABYSS program, as well as the largest overlap in targets between them. Note that due to a large number of cartons, overlap between some of the cartons cannot be shown.
\label{fig:chart}}
\end{figure}

\begin{figure*}
\epsscale{1.0}
\plotone{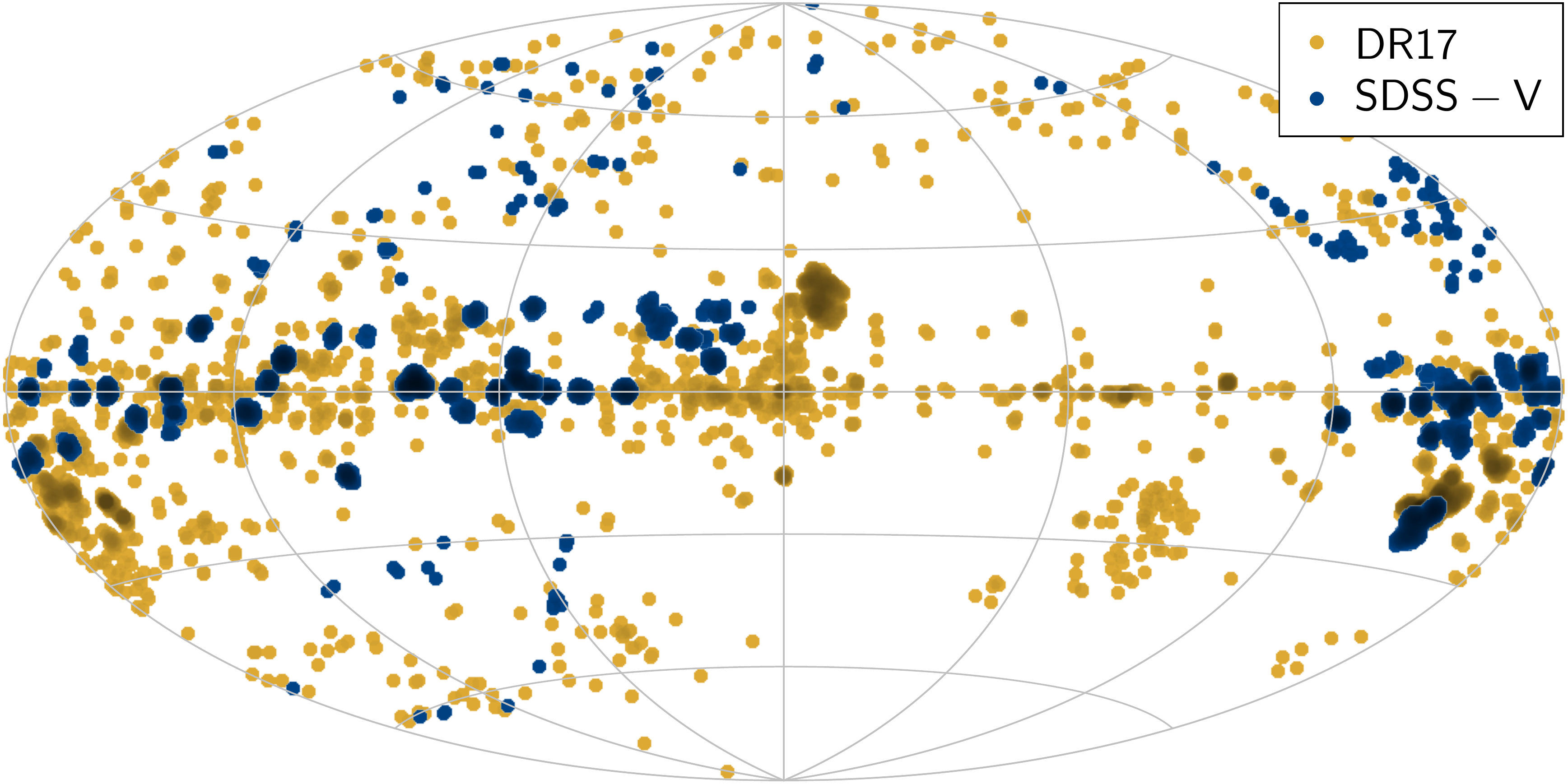}
\caption{Distribution of ABYSS sources that have been observed with SDSS through 2021. Stars in yellow have been observed with APOGEE prior to the beginning of SDSS-V. Sources in blue are those that have been observed as a part of SDSS-V plate program, typically with both APOGEE and BOSS fibers assigned in a given field.
\label{fig:obs}}
\end{figure*}

\subsubsection{PMS}

There have been several dedicated studies that focused on identifying pre-main sequence stars using an HR diagram generated through Gaia photometry and astrometry. We incorporate the resulting catalogs from two such works.

\citet{zari2018} have selected low-mass pre-main sequence stars using Gaia DR2 data within 500 pc that are found above (and therefore are younger than) the 20 Myr PARSEC isochrone \citep{marigo2017} and fainter than $M_G>4$ mag. The photometry has been first extinction corrected, excluding sources with $A_G<0.92$ mag, to avoid reddened field stars. Furthermore, additional cuts have been made to select stars with low parallax errors ($\sigma_\pi/\pi<20$\%), to limit the sample to disc stars (total tangential velocity $\sqrt{v_l^2+v_b^2}<40$ \kms), and to produce a "clean" HR diagram as suggested by \citet[][Appendix C]{lindegren2018}.

This selection is effective for nearby populations, but, at larger distances the sample becomes strongly contaminated by field stars, both due to an imperfect match of real photometry to the isochrones, and due to imperfections in the extinction correction. An alternative approach was considered by \citet{mcbride2021}, using a neural network Sagitta. It was trained on Gaia and 2MASS photometry of stars in young populations from \citet{kounkel2020} to autonomously identify low-mass pre-main sequence stars, automatically adjusting color-magnitude threshold based on age and distance of a star. The constructed sample extended up to $\pi>0.2$ mas. It consisted of sources with classification probability $>70$\%, and it also had several data quality criteria, such as $\sigma_\pi/\pi<10$\% or $\sigma_\pi<$0.1 mas, precision in Gaia photometry in all bands $<10$\%, recommended cuts based on the photometric excess noise, as well as Gaia RUWE$<$1.4.

These two catalogs form the basis of YSO\_PMS\_APOGEE and YSO\_PMS\_BOSS cartons that were first introduced in V0.5 version of targeting, containing 76,332 and 73,213 stars respectively. Initially, both catalogs used Gaia DR2 data; in V1 the catalog from \citet{mcbride2021} was upgraded to EDR3 version which (due to magnitude limits) did not change substantially, except for a minor improvements in sensitivity to more distant PMS stars.

\subsubsection{OB}
Most of the YSO cartons (with exception of Cluster and CMZ) focus exclusively on low-mass YSOs, as they are most distinct from field stars. Intermediate and massive stars, on the other hand, reach the main sequence very quickly, and thus become difficult to differentiate. 

However, as OB stars have short lifetimes, they would always be young. Thus, to fill the gap in targeting, we selected sources based on examining the placement of known OB stars from \citet{maiz-apellaniz2016}.

\begin{itemize}
\item $-0.2<(G_{BP}-G_{RP})<1.1$
\item $M_G<1.6(G_{BP}-G_{RP})-2.2$
\item $G<18$ mag, $\pi>0.3$ mas
\end{itemize}

In V0, this selection formed the basis of YSO\_OB carton, which was split into YSO\_OB\_APOGEE and YSO\_OB\_BOSS, both containing 8,670 stars. However, as all of the selected stars are very bright (typically $G<12$ mag), they currently cannot be observed with BOSS without offsetting their positions, due to the saturation limit of the instrument.

In V0.5, these cartons were rendered obsolete, as all of the targets that are a part of YSO\_OB are a perfect subset of OBA\_CORE program within SDSS-V \citep{zari2021}, thus, they do not require duplication of efforts from multiple programs.

\subsection{Sample summary}

The map showing the spatial distribution of all stars in all cartons is shown in Figure \ref{fig:map}. Unsurprisingly, almost all sources are found along the Galactic plane and Gould's belt. The angular scale height of the disk between the cartons strongly depends on the typical distance of the stars within it. Of the optically bright sources, YSOs in the vicinity of the Solar Neighborhood dominate the PMS carton, while sources found beyond $>$1 kpc dominate the OB carton (Figure \ref{fig:dist}).

In total, V0.5 targeting sample consists of 202,726 sources to be observed with APOGEE, and 196,188 sources to be observed with BOSS. There is some overlap between the cartons, but, in general, most of the sources in each carton are unique, sensitive to distinct tracers of youth (Figure \ref{fig:chart}). For example, only 10\% of stars in Disk carton are also found in PMS carton. This is partially due to PMS carton requiring high precision photometery and astrometry, which can often be poor in stars with disks, even in the optically bright stars. Furthermore, PMS carton is primarily sensitive to the nearby stars, whereas Disk carton can include many more distant stars along the plane of the disk. Similarly, only 25\% of stars in Cluster carton are also found in PMS carton. Clustering is unbiased to \teff, as such Cluster carton includes many more high mass stars than what can be selected as high fidelity PMS star based on photometry alone. On the other hand, clustering preferentially selects regions with high stellar density, and it often struggles recovering more diffuse groups, such as, e.g. Taurus, or outer parts of the populations with a strong density gradient, or older groups that are starting to lose dynamical coherence. As such, all of the targeting approaches are highly complementary to one another.

The catalog of sources is available as a part of SDSS DR18 \citep{almeida2023}. It can be accessed through SkyServer\footnote{\url{https://skyserver.sdss.org/dr18/SearchTools/sql}} using an ADQL query:
\begin{verbatim}
SELECT TOP 1000
mc.carton,mt.ra,mt.dec
FROM mos_carton mc
JOIN mos_carton_to_target mctt ON
    mc.carton_pk=mctt.carton_pk
JOIN mos_target mt ON
    mt.target_pk=mctt.target_pk
WHERE CHARINDEX("yso", mc.carton) > 0
\end{verbatim}
This will return the first 1000 sources; an individual source would be included multiple times for every single carton in which it appears.

\section{First year data}\label{sec:firstyear}

\begin{figure*}
\epsscale{1.0}
		\gridline{\fig{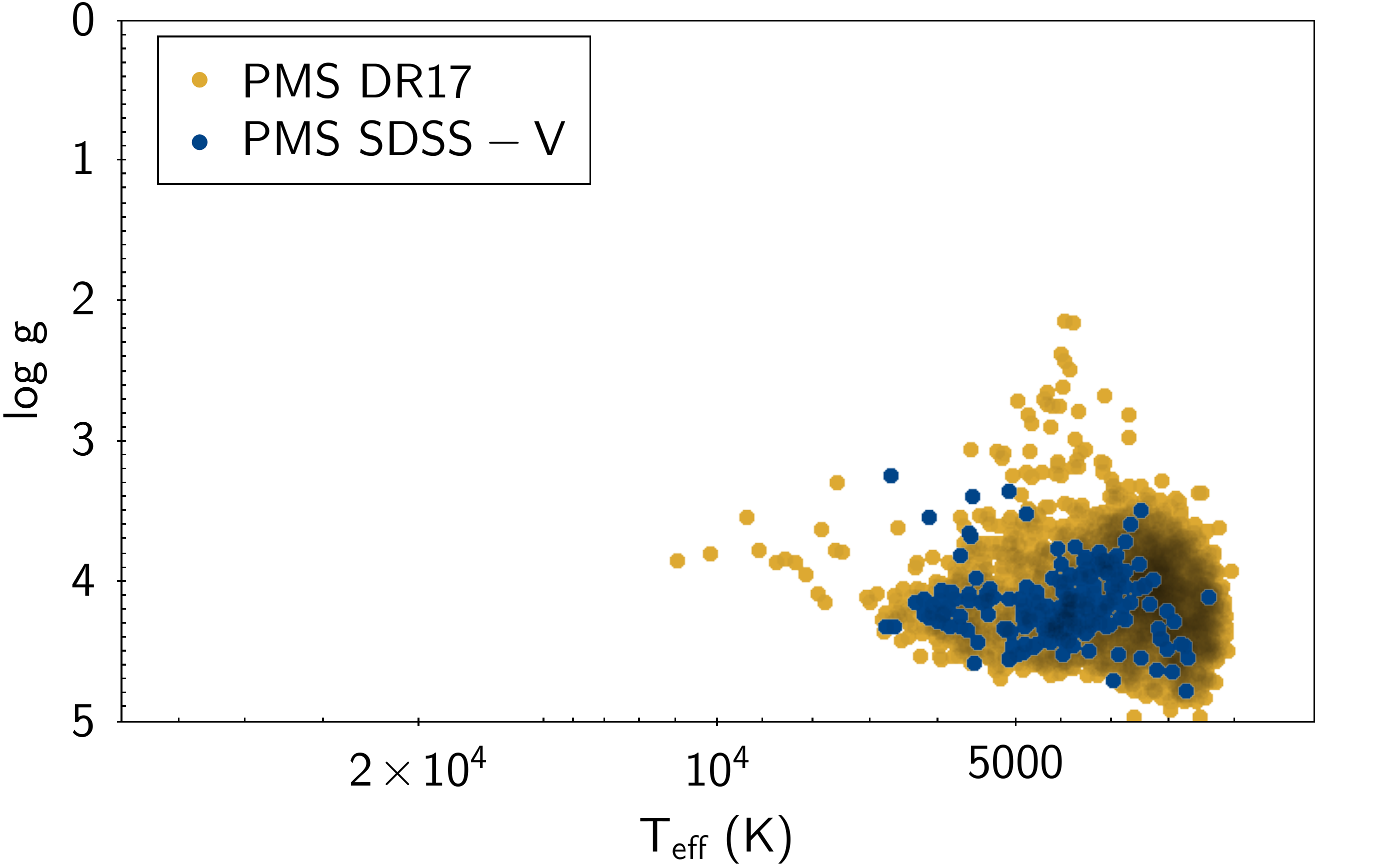}{0.5\textwidth}{}
		          \fig{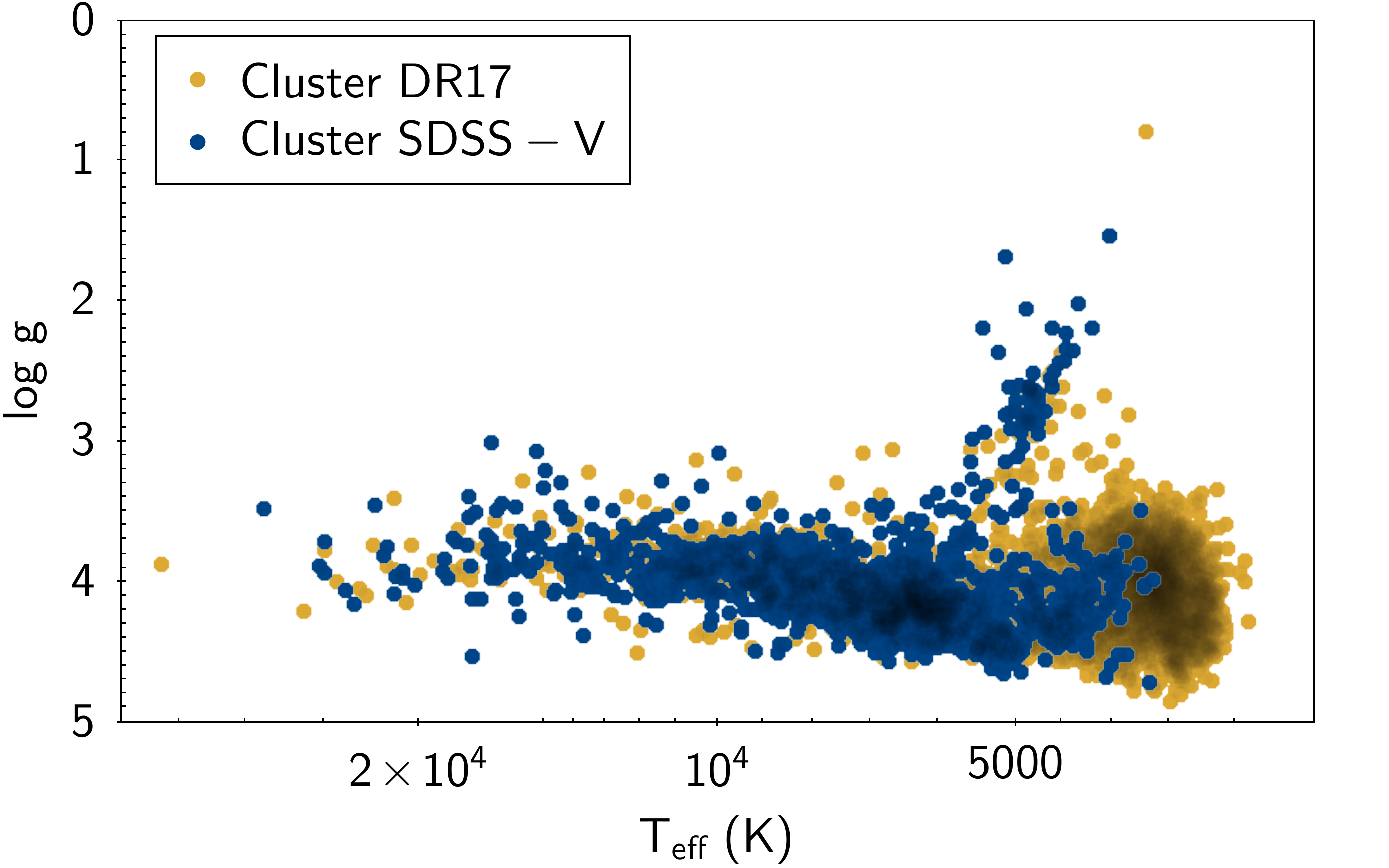}{0.5\textwidth}{}
		 }\vspace{-1cm}
		\gridline{\fig{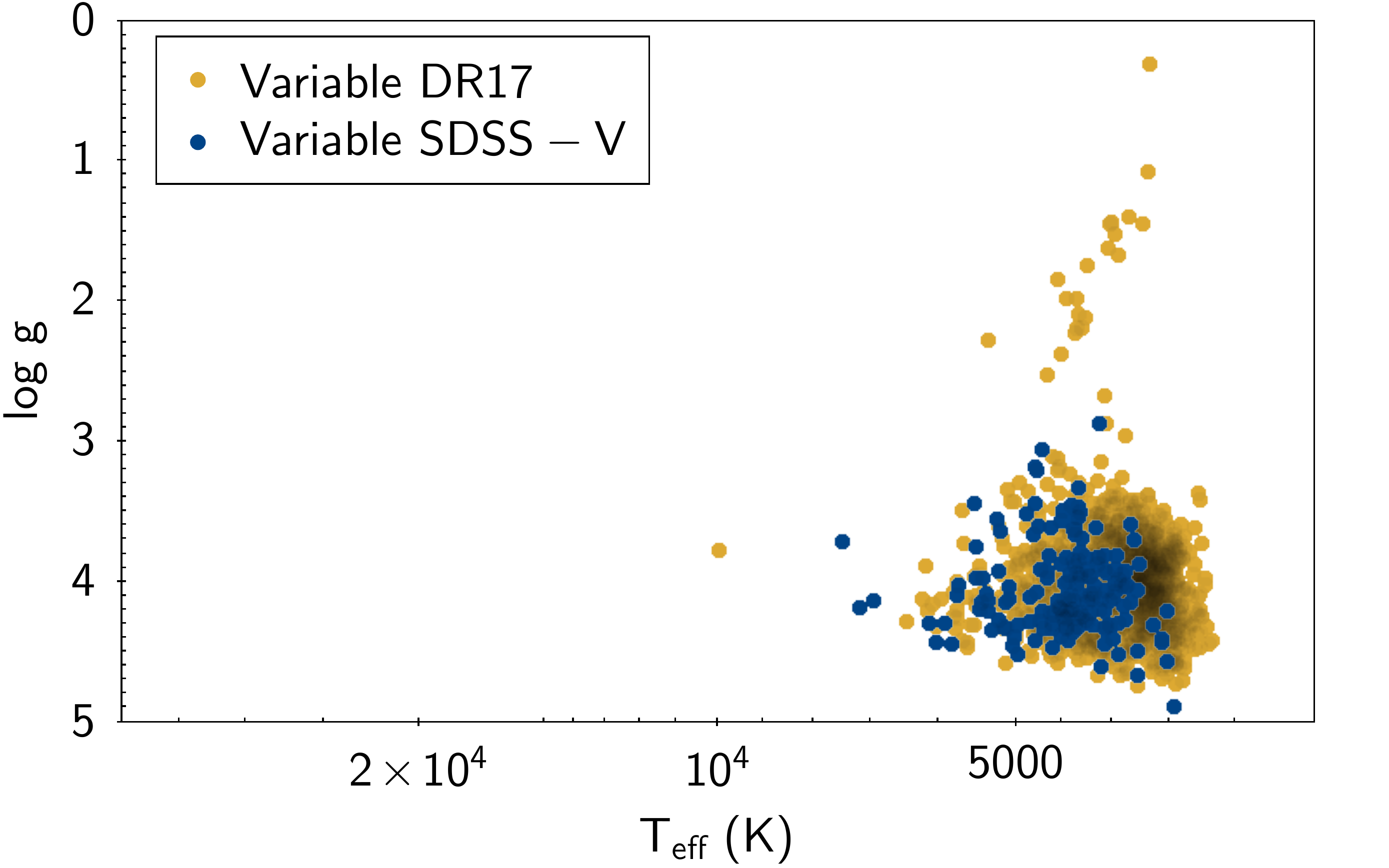}{0.5\textwidth}{}
		          \fig{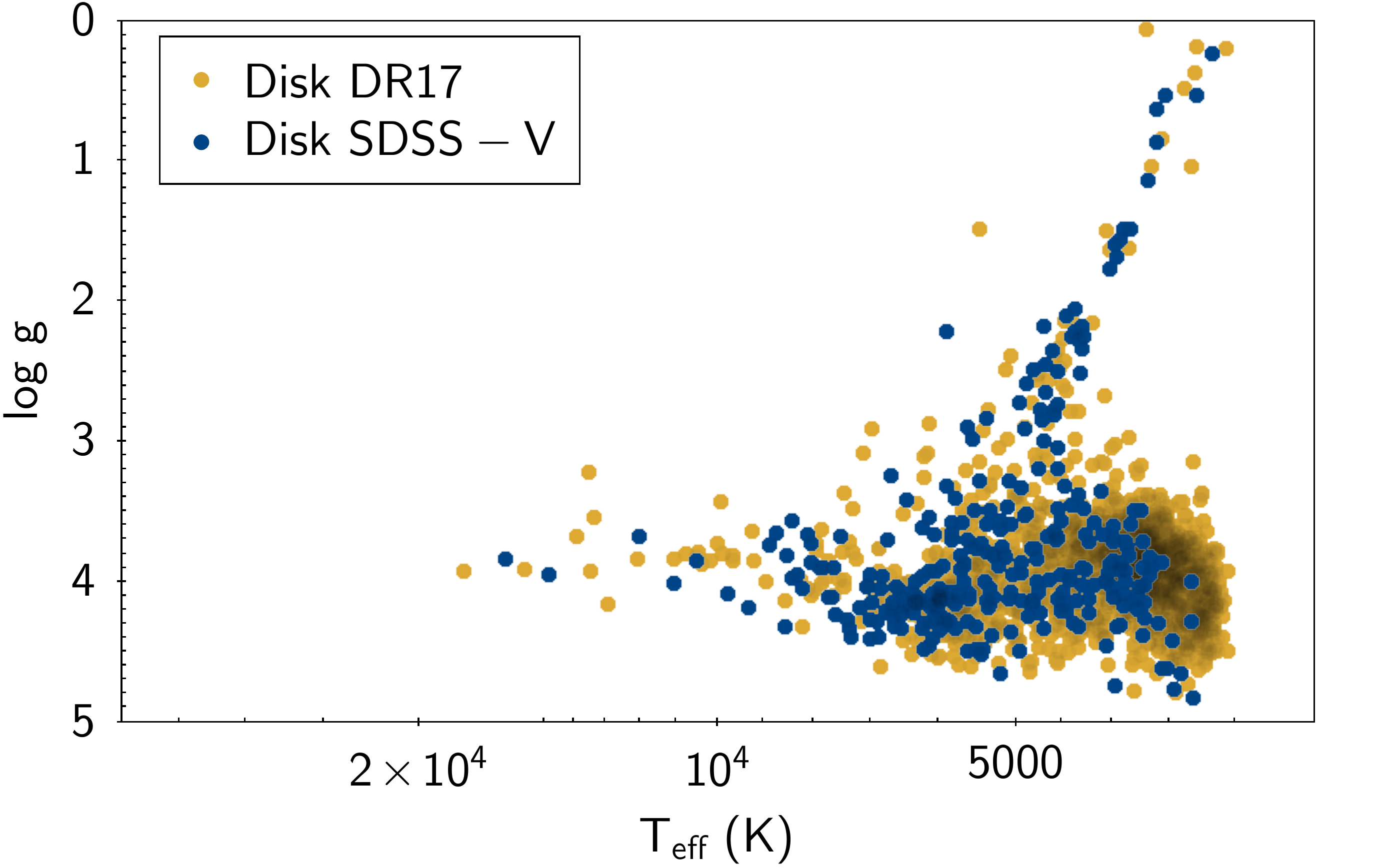}{0.5\textwidth}{}
		 }\vspace{-1cm}
		\gridline{\fig{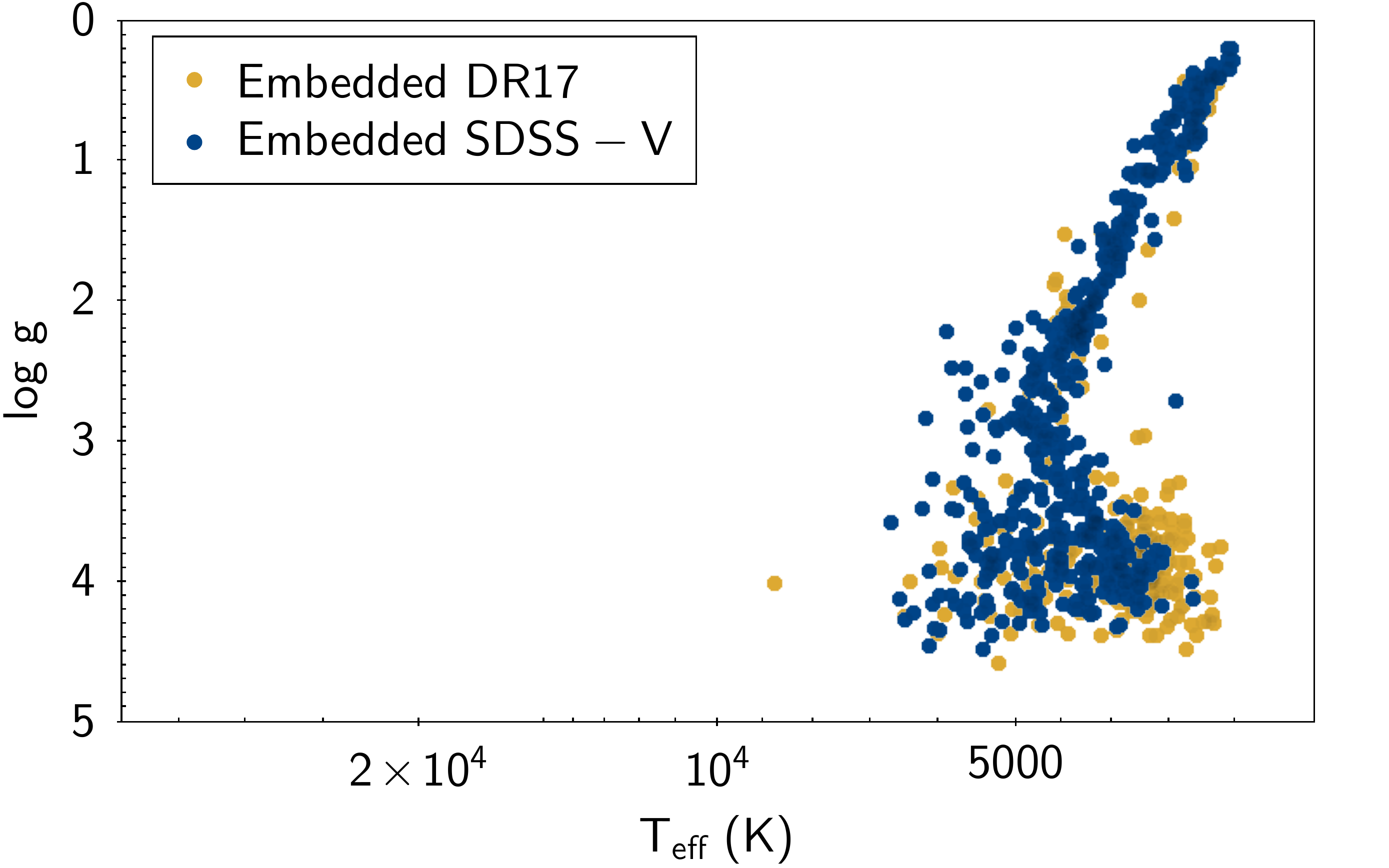}{0.5\textwidth}{}
		          \fig{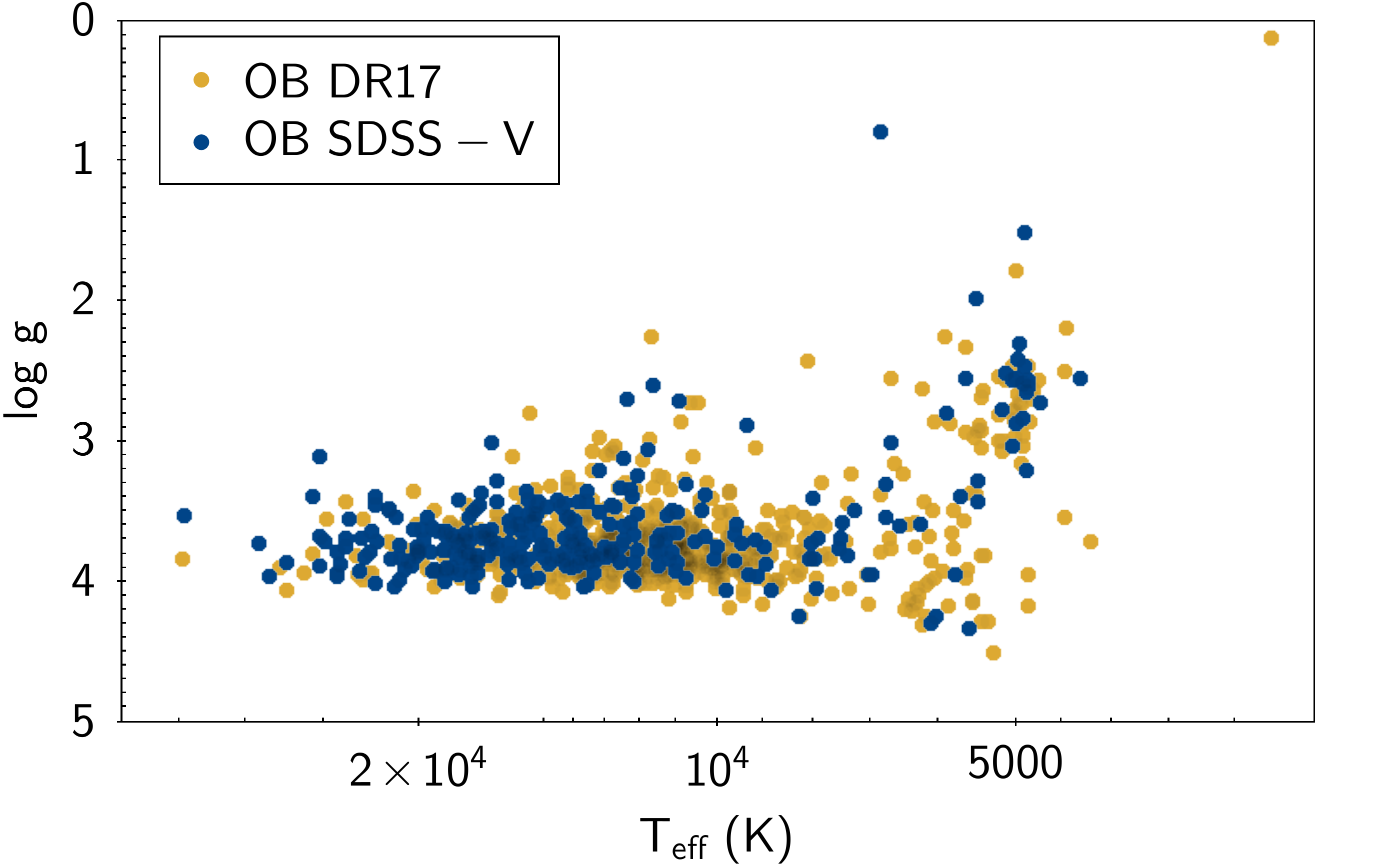}{0.5\textwidth}{}
		 }\vspace{-1cm}
		\gridline{\fig{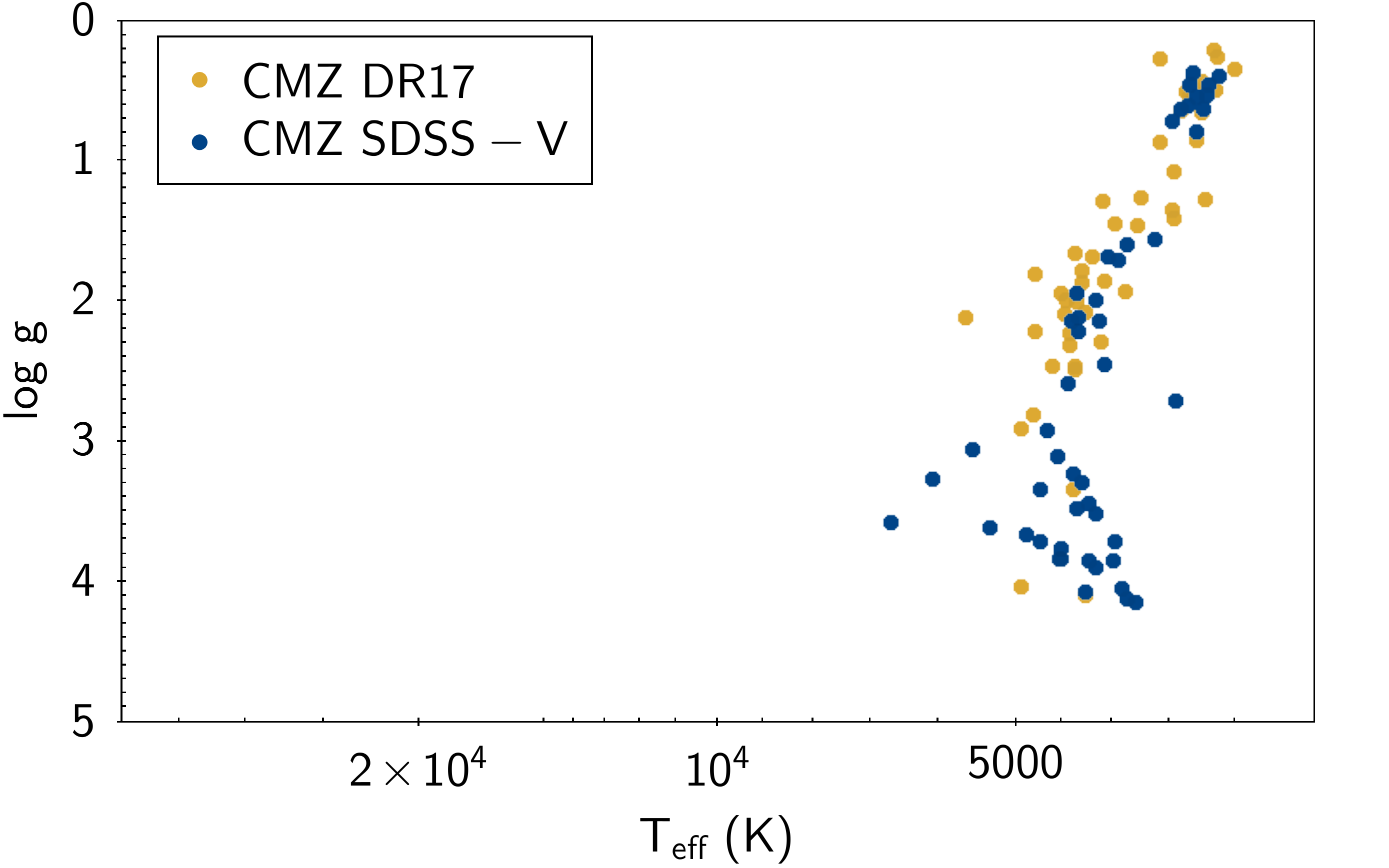}{0.5\textwidth}{}
		          \fig{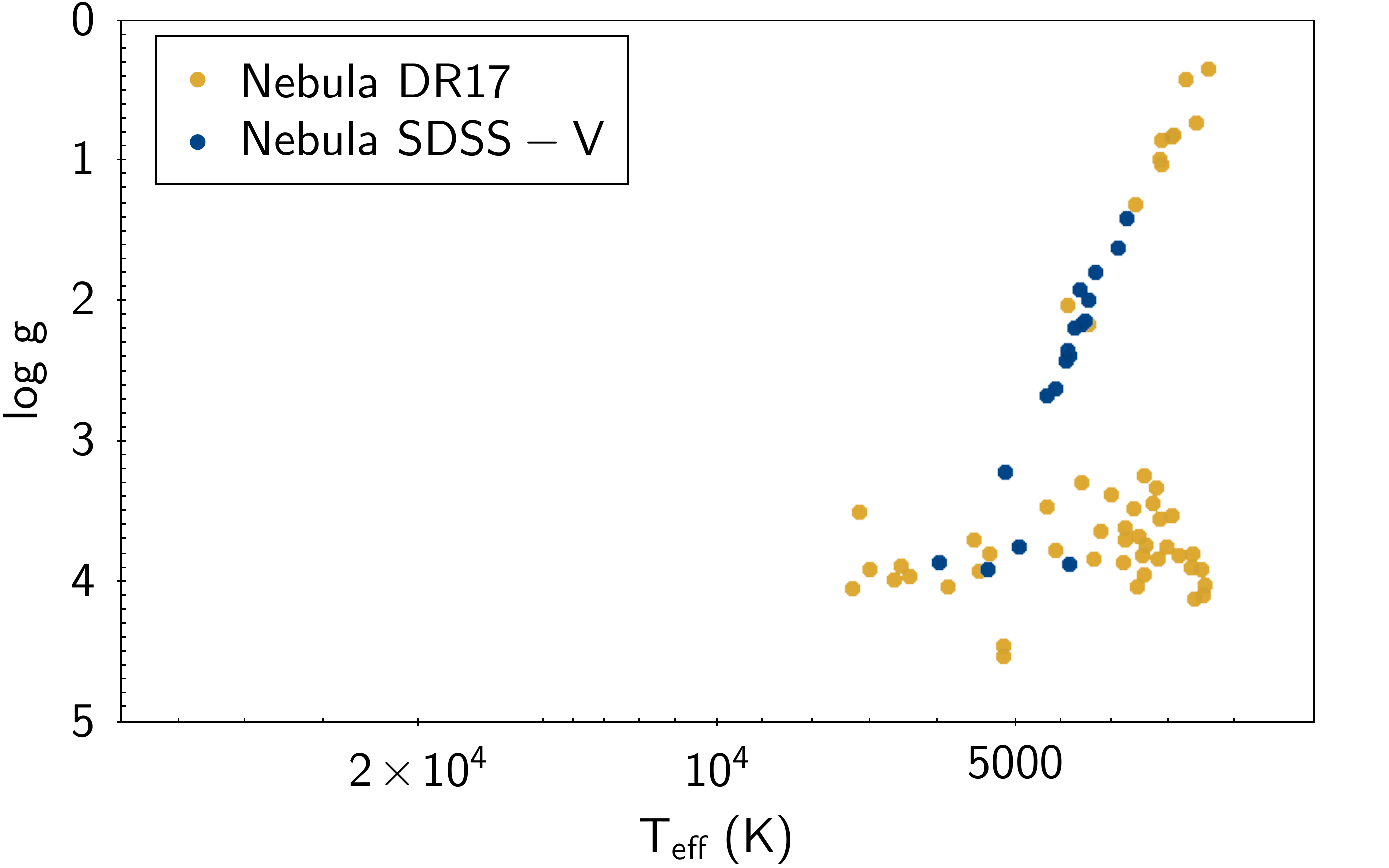}{0.5\textwidth}{}
		 }\vspace{-0.5cm}
\caption{Spectroscopic parameters extracted by APOGEE Net from the APOGEE spectra for new and archival ABYSS targets, highlighting the typical temperature range encompassed by each carton, as well as a typical rate of contamination. Note that most sources occupy the parameter space expected by the pre-main sequence stars, with only minor contamination from the red giants in most cartons.
\label{fig:anet}}
\end{figure*}

In this section we describe the data that have been obtained in the course of the first few months of SDSS-V operations as well as the available archival data for the ABYSS targets. We give an overview of the data processing pipelines that are currently available to process these data, and their fidelity.

\subsection{Planning of observations}
Although SDSS-V began its operations in 2021, it did not immediately reach its full capability. As the transition to the robotic fiber positioning system (FPS) in both hemispheres required significant instrument upgrades, initial observations at APO were carried out with the SDSS-IV plug-in plate system \citep{wilson2019} during the first six months of operations. As young stars can often be found inside compact clusters, and the plug plates can pack the fibers closer together than the robot positioners, such a configuration was deemed advantageous for ABYSS. Thus, several plates were commissioned to target fields with the highest density of sources that had not been targeted in previous iterations of the survey. Additionally, YSO targets were included in the 2021 plates led by other programs.

In the previous iterations of the survey, BOSS was primarily used in dark time to look at faint targets. As such, to avoid saturation for a 15-minute exposure, the nominal bright limit for BOSS is $G>13$ mag, whereas for APOGEE is $H>7$ mag. In the near future, such limits will be overcome through offsetting the fiber position from the position of a star. Meanwhile, for the first year plate operations it was decided that ABYSS targets would be split: sources brighter than $G_{RP}<13$ mag would be observed with APOGEE, and fainter stars would be observed with BOSS. At the moment, the split is sub-optimal, due to a resulting segregation in mass. However, once instrumental limitations are lifted, the survey is expected to obtain a complementary set of observations to compensate this problem.

The SDSS-V plate program has obtained spectra for 2,462 ABYSS targets with APOGEE, and 4,854 targets with BOSS. These data can be supplemented by 7,884 sources for which archival APOGEE spectra exists as part of DR17 \citep{abdurrouf2022}. These include young stars targeted explicitly by SDSS III \& IV, and those that have been serendipitously targeted by other programs (Figure \ref{fig:obs}). In these data, 193 sources have both APOGEE and BOSS spectra.

\subsection{APOGEE}

Due to the extensive history of observations, various pipelines have been developed to measure stellar parameters (such as \teff\ and \logg) from APOGEE spectra \citep{cottaar2014,olney2020,sprague2022}. The latest iteration of these pipelines, APOGEE Net, can extract parameters for all stars with \teff$>3000$ K in a self-consistent manner, and for PMS stars its \logg\ is sensitive to age. Thus, using the latest iteration of the APOGEE Net pipeline, we qualitatively evaluate how prone is each individual carton to field contamination from red giants (Figure \ref{fig:anet}). Full quantitative assessment that also considers contamination from the older main sequence stars will be presented in future papers in this series.

Optically bright cartons (PMS, Variable, Cluster, Disk, and OB) have only minimal contamination from red giants ($\sim$4\%) relative to the total number of sources. Approximately half of the sources in the Embedded carton are red giants. The bulk of this contamination can be reduced in the future through the 2MASS color cut introduced in V1 version of targeting (Figure \ref{fig:dusty}). The contamination rate for other two optically faint cartons, CMZ and Nebula, is difficult to evaluate at the moment due to the limited size of the samples, as existing SDSS-V observations do not cover the region of space where the bulk of these targets reside. Similarly, archival DR17 observations impose their own targeting selections that may favor a particular type of sources, as such they may not necessarily be representative of a full set of sources for these cartons. E.g., as during SDSS-III and -IV, the primary targets for the survey were stars that have photometry consistent with being red giants, which may make the red giant contaminants to be over-represented in the sample observed prior to SDSS-V. 

Different cartons favor different temperature ranges. The PMS and Variable cartons typically select sub-solar systems: they are designed for low-mass stars that are convective and are slow to reach the main sequence. Disk and Embedded cartons also favor low-mass stars: the Embedded carton is likely to have fewer massive stars than the Disk carton, as higher mass stars can remain optically bright through a higher degree of extinction. Nonetheless, low-mass stars have longer disk lifetimes \citep[e.g.,][]{bayo2012,ribas2015}. Sources in the Cluster carton are not based on stellar properties of individual stars, but rather on their membership, as such stars of all masses are represented in it. Finally, sources in OB carton favor sources with \teff$>$10,000 K. In comparison to massive stars in the Cluster carton, sources in the OB carton tend to skew towards hotter \teff\ and lower \logg\ (on average by $\sim0.1$ dex).

Typically the lowest \teff\ observed in the DR17 data is lower than is currently available for SDSS-V data. As mentioned previously, this is due to preferentially observing brighter stars with APOGEE due to the saturation limit in BOSS.

\subsection{BOSS}

\begin{figure*}
\epsscale{1.0}
\plotone{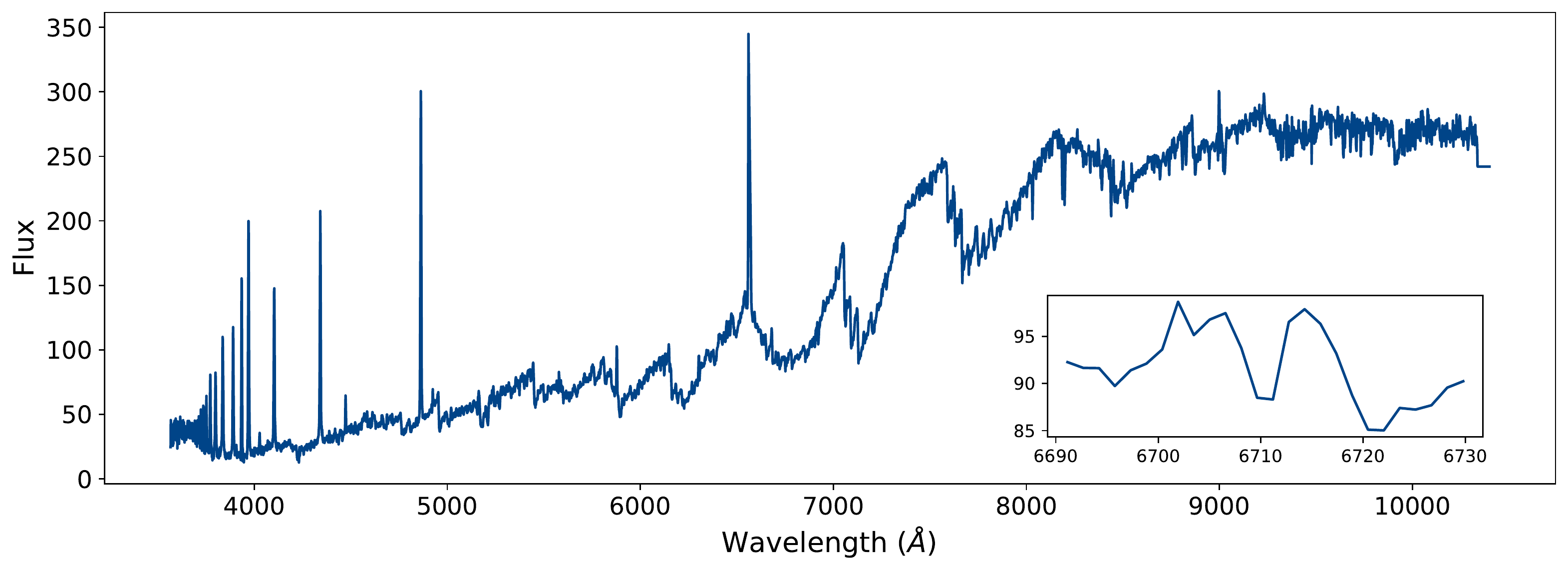}
\caption{An example of a BOSS spectrum of a young star. The inlay shows the Li I 6707.7 \AA\ line.
\label{fig:boss}}
\end{figure*}

\begin{figure*}
\epsscale{1.0}
    \includegraphics[width=0.53\linewidth,trim=15 0 0 0,clip]{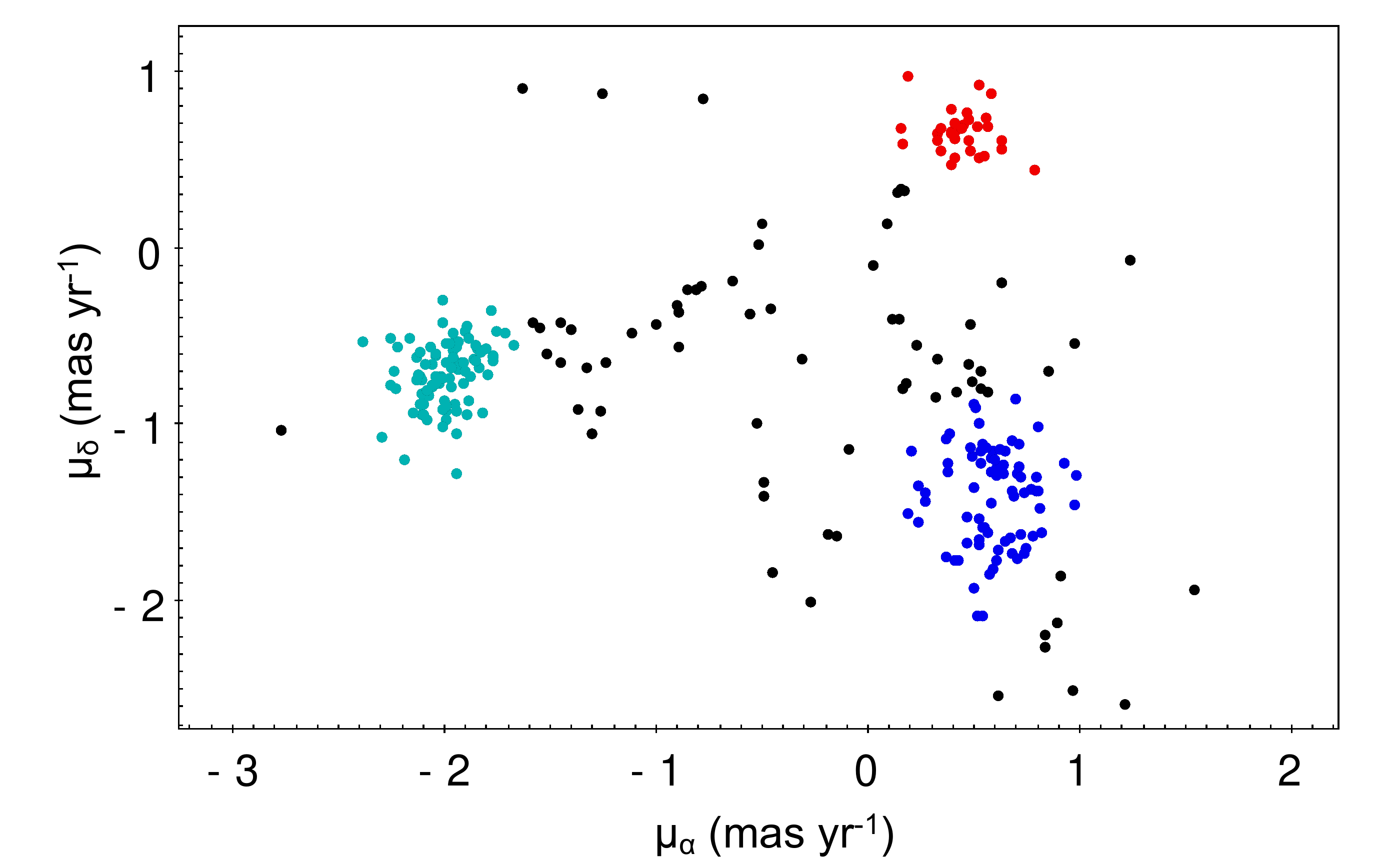}
    \includegraphics[width=0.47\linewidth]{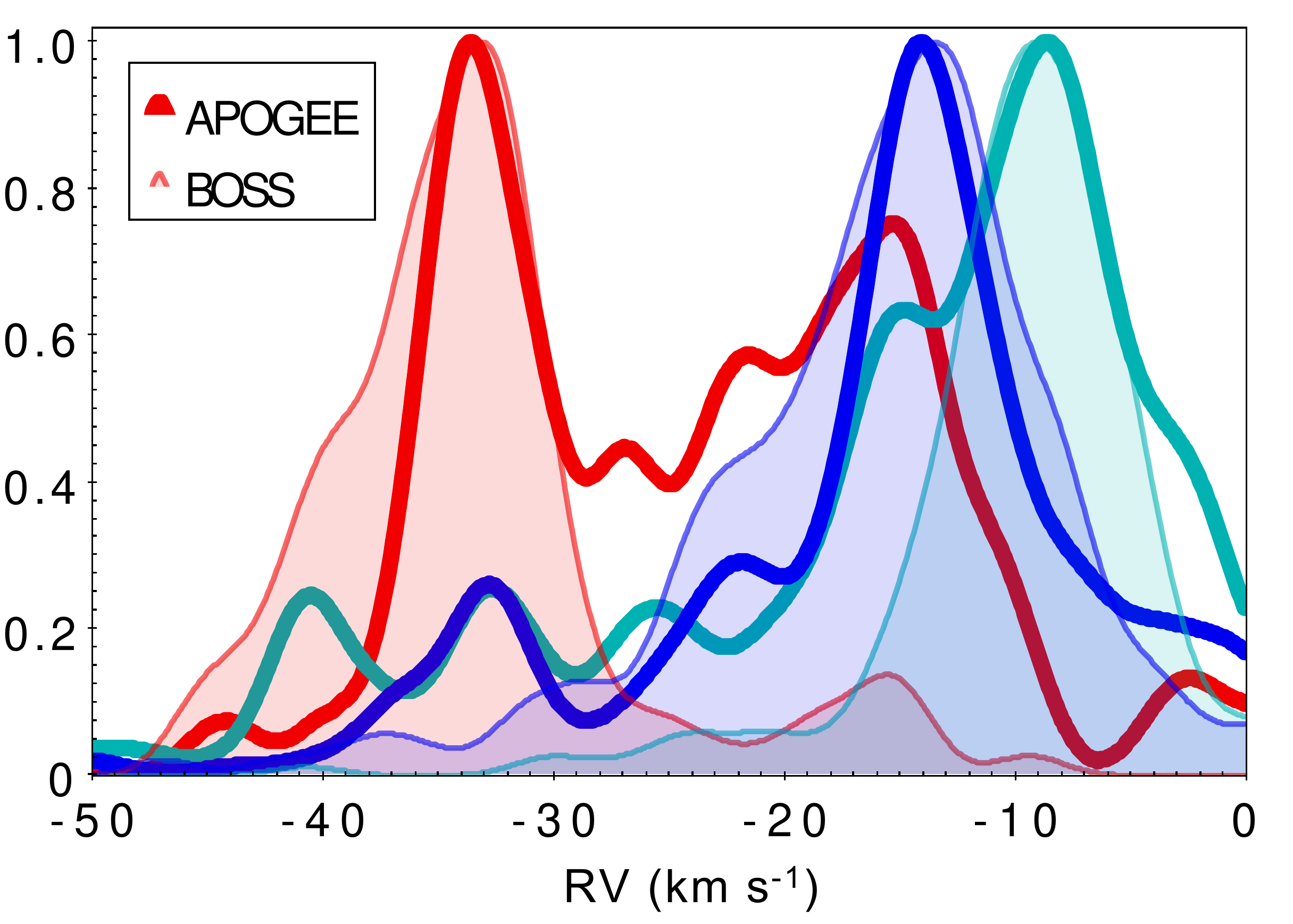}
\caption{Velocity structure observed towards a field observed by ABYSS towards Cam OB1 association. Three kinematically coherent groups are located towards it, all found in a similar location on the sky, and at a similar distance. These groups are distinguishable in the proper motion space (left), and in RV space (right, same colors). The same velocity structure is recovered both by APOGEE (thick unshaded curve) and BOSS (shaded curve).
    \label{fig:comissioning}}
\end{figure*}

YSOs have a long history of being observed with APOGEE. On the other hand, fewer than 40 ABYSS targets have archival BOSS observations, most of them concentrated near 25 Ori \citep{suarez2017}. LAMOST (Large Sky Area Multi-Object Fiber Spectroscopic Telescope) low resolution spectroscopy is comparable to BOSS, both in wavelength coverage and spectral resolution. In LAMOST DR7, spectra are available for 13,452 ABYSS targets, including several prominent star forming regions. Despite this, to date only a few studies of YSOs utilized LAMOST spectra \citep[e.g.,][Hernandez, J. et al. in prep]{liu2021,wang2022c}. 

As such, with the exclusion of radial velocities (RVs), currently there are no pipelines capable of extracting reliable stellar parameters from either BOSS or LAMOST spectra of low-mass YSOs. There are however existing efforts to rectify this, both by measuring reliable \teff\ \& \logg\ (e.g., Sizemore, L. et al. in prep), as well as  characterizing a number of features found in the spectral range of the instrument that could be used as indicators of youth (such as Li I, or various emission lines for H or Ca; Saad, S., et al. in prep; Figure \ref{fig:boss}).

At the moment, BOSS spectra are processed by pyXCSAO \citep{pyxcsao}, which is a Python implementation of IRAF RVSAO package \citep{tonry1979,rvsao}. The spectra are cross-correlated against synthetic PHOENIX templates \citep{husser2013}. Subgrid solutions are derived for parameters such as \teff\ and \logg\ using the quality of the fit. However, there are significant systematics that affect the quality of the derived parameters for the young stars in particular, not dissimilar to what was observed in the original APOGEE YSO pipeline \citep{cottaar2014}. RVs are measured from the best fitting template for stars with \teff$>$3,500 K. RVs for cooler stars are determined from best fitting 3,500 K template, as they otherwise appear to be systematically redshifted relative to the rest velocity of other stars in the same star forming region. This has also been seen in the APOGEE data, \citep[e.g.,][]{kounkel2019}, due to a systematic issue in synthetic spectra of cool stars.

\begin{figure}
\epsscale{1.2}
\plotone{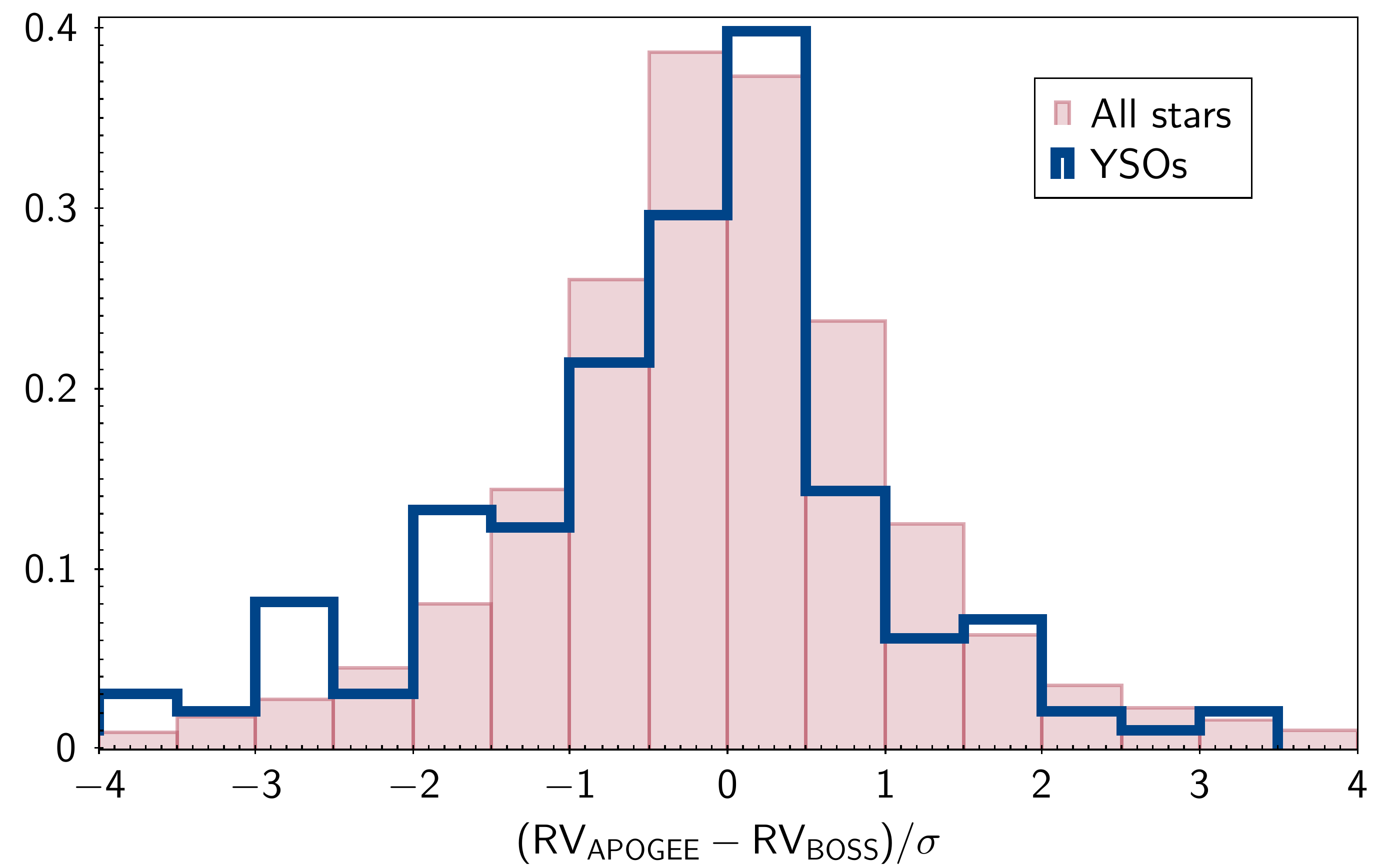}
\caption{Difference between APOGEE and BOSS RVs for the same stars, divided by BOSS uncertainties in the RVs. All stars, regardless of the evolutionary status, observed to date are shown in red. ABYSS-only targets are shown in blue. Note that the typical scatter is consistent within 1$\sigma$; wider wings may be attributable to spectroscopic binaries.
\label{fig:rvdiff}}
\end{figure}

APOGEE can achieve sub-\kms\ precision in its measured RVs. BOSS, being a lower resolution instrument, can produce RV precision of only $\sim$4--5 \kms\ for spectra of low-mass stars with high signal-to-noise. Nonetheless, both instruments have been vetted to ensure consistent performance and a lack of a zero-point offset between them, both in the average properties derived for individual regions (Figure \ref{fig:comissioning}), and in the direct comparison of RVs of the individual stars, when possible (Figure \ref{fig:rvdiff}).

We defer a more detailed analysis of the BOSS spectra and the parameters to subsequent publications in the series.

\section{Gaia DR3 comparison}

\begin{figure}
\epsscale{1.2}
\plotone{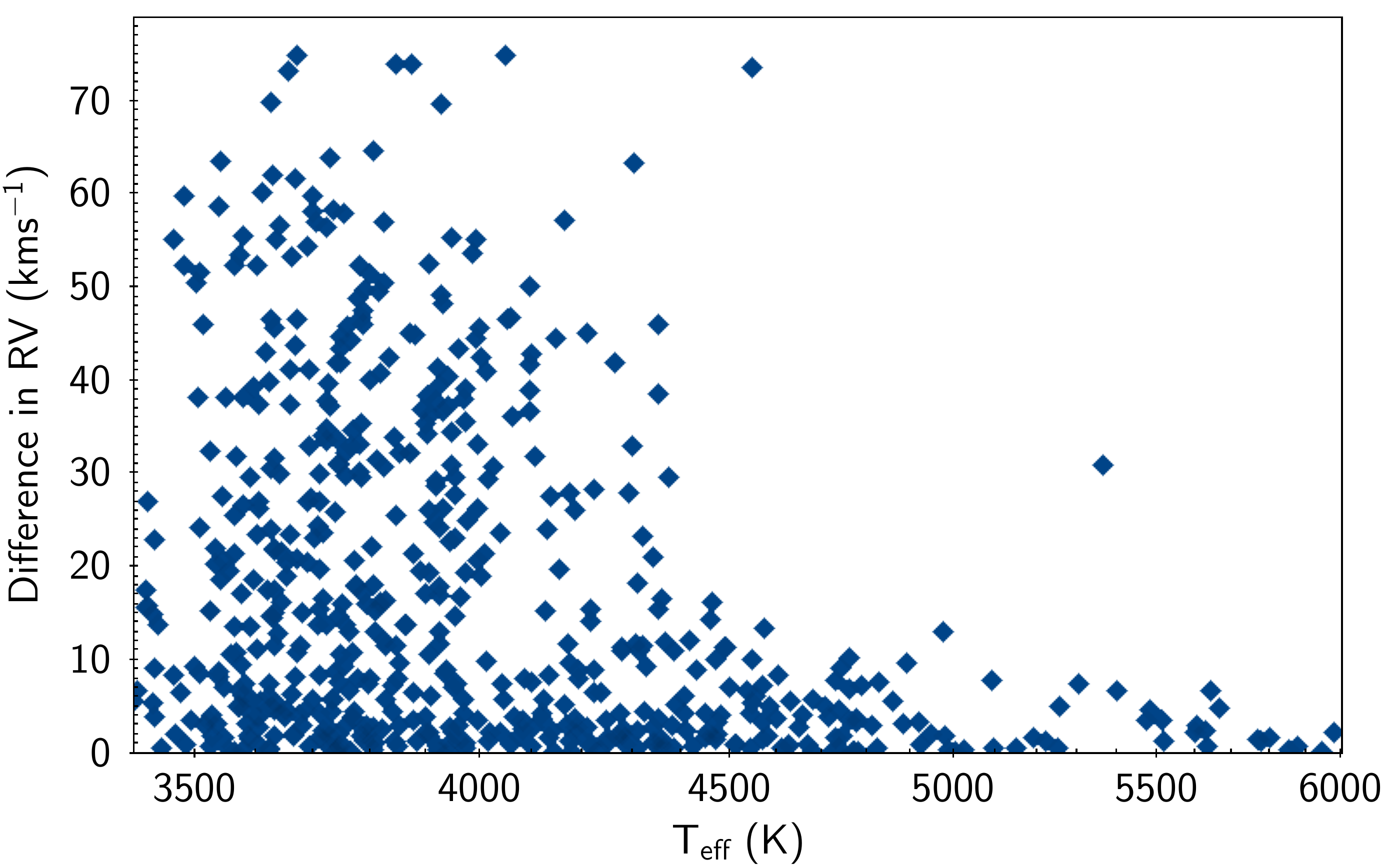}
\caption{A comparison of RVs for young stars between APOGEE and Gaia DR3. Sources have been selected from \citet{kounkel2019} for which at least 3 APOGEE epochs have been obtained to confirm their RV stability, excluding any of the spectroscopic binaries. The typical precision in RV is $<$1 \kms\ for APOGEE, and $\sim$6 \kms\ for Gaia.
\label{fig:gaiacomp1}}
\end{figure}
\begin{figure*}
\epsscale{1.1}
\plottwo{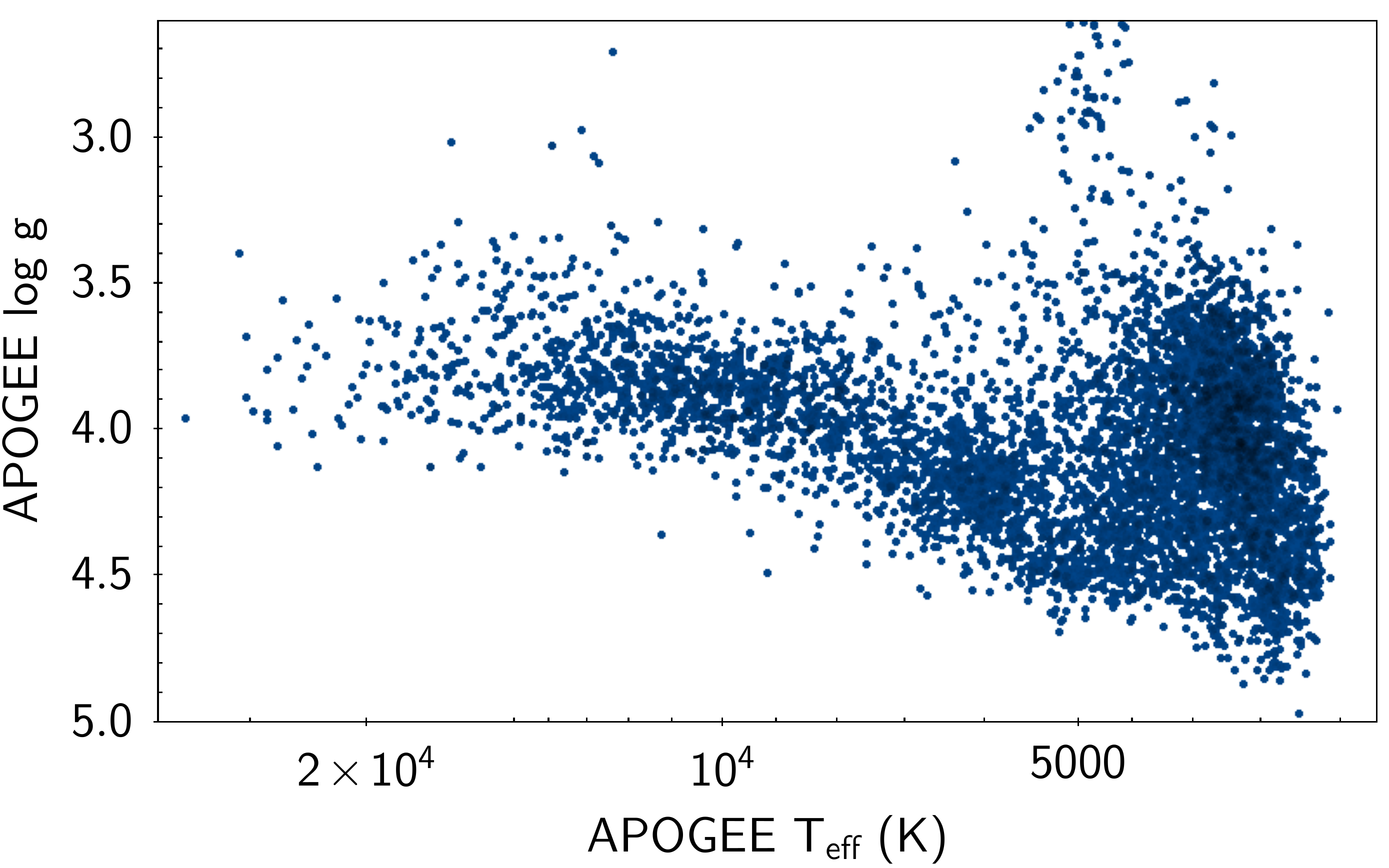}{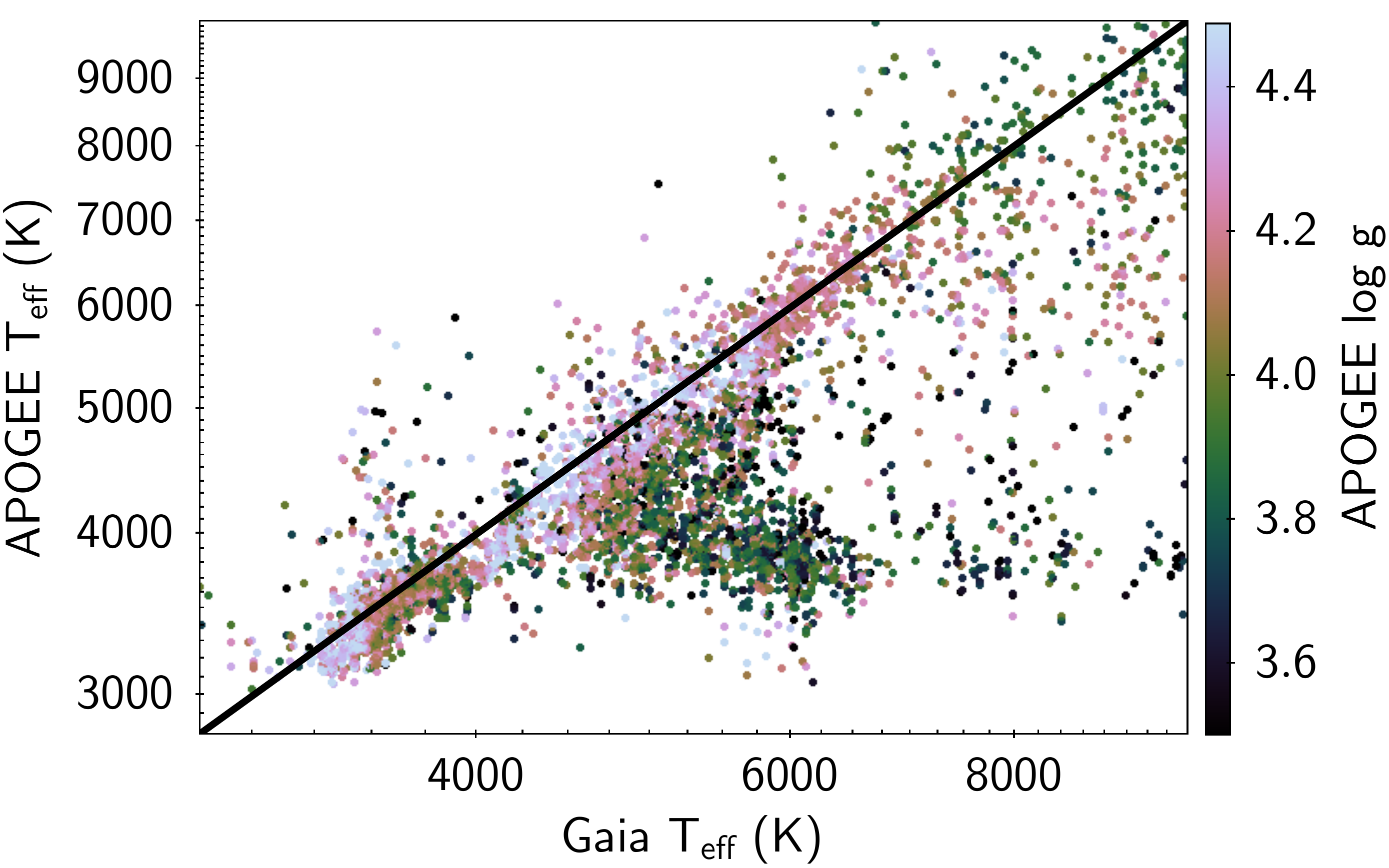}
\plottwo{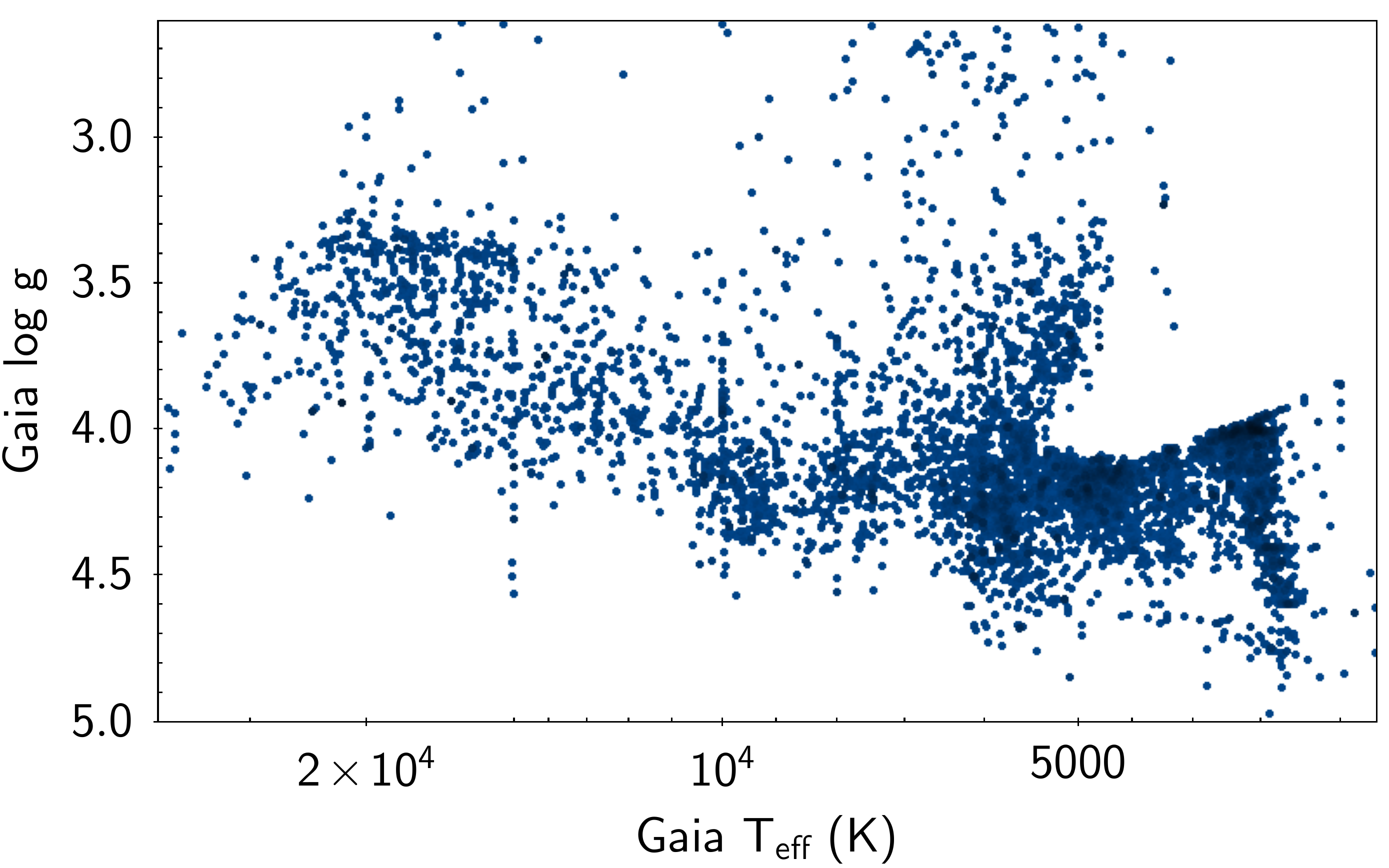}{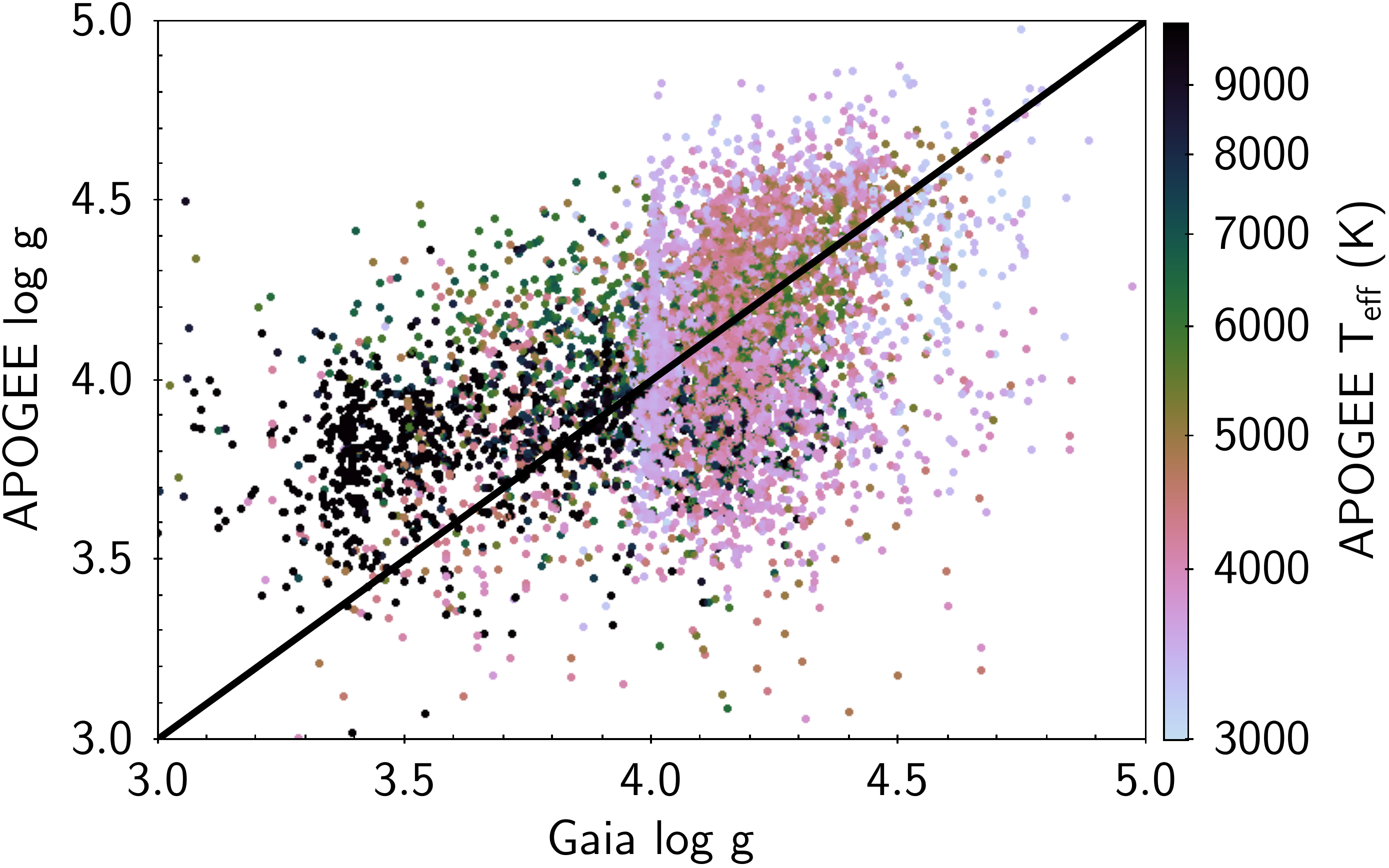}
\caption{A comparison of APOGEE-derived \teff\ \& \logg\ for the ABYSS objects, vs those in Gaia DR3.
\label{fig:gaiacomp}}
\end{figure*}
The recent third Gaia data release \citep[Gaia DR3,][]{gaia-collaboration2022a} has made available not just the astrometric parameters that have been present in Gaia DR2 or EDR3, but also RVs for 30 million stars derived with its on-board spectrograph \citep{katz2022}, as well as stellar parameters such as \teff\ and \logg\ for 5.5 million stars \citep{fouesneau2022}. This census includes a number of ABYSS-targeted objects.

Unfortunately, these parameters were optimized to produce accurate solutions for ``typical'' stars. YSOs do not fall into this category, due to a number of unique spectral features they exhibit from both accretion and activity, especially near Ca II triplet at $\sim$8500 \AA, which is tightly encompassed in the spectral range of Gaia RVS spectrograph \citep{recio-blanco2022}. Thus, caution has to be utilized in interpreting the available parameters in this data release for the young stars. 

In particular, \citet{kounkel2022c} identified issues in Gaia RVs for YSOs: 1) they are not precise, as they have typical uncertainties of 5--10 \kms\, which is significantly worse than the instrumental limit at the typical signal-to-noise of these observations. 2) they are not accurate as when evaluated against high resolution RVs from existing APOGEE observations of spectroscopically stable young stars, they typically show a scatter $>5\sigma$ (Figure \ref{fig:gaiacomp1}). This issue persists for more than 100 Myr, as a large scatter in RVs among low-mass stars is detected in a (comparatively) older cluster such as the Pleiades.

Similarly, we evaluate \teff\ and \logg\ of the ABYSS stars that Gaia has observed, and compare them against the derived parameters from APOGEE (Figure \ref{fig:gaiacomp}). We find that  the parameter space occupied by cool pre-main sequence stars (\teff$<$5000 K, low \logg) is currently not well sampled by the pipelines employed by the \textit{Gaia} consortium. The sources that occupy this parameter space are often recognized to have  lower \logg\ than what is commonly found in main sequence stars, but they are pushed towards hotter \teff, which results in placing them towards the red giant branch. No correlation is found between the reported \logg\ values for low mass stars between these two data sets, however. Further, the Gaia-derived \logg\ are not sensitive to stellar ages, unlike those from APOGEE \citep[e.g.,][]{olney2020,kounkel2022b}. 
Similar caution should be given to other derived parameters produced by Gaia for these stars, such as, for instance ages.

We stress that this issue is specific to young stars, most prominently on the low-mass end, and should not affect more evolved main sequence stars that are older than a few 100s Myr. In future releases, the Gaia spectra may be reprocessed with a pipeline that is better tuned to YSOs, both within the collaboration or through efforts using the publicly released spectra. However, as Gaia DR4 is not expected until 2026 at the earliest, by which point ABYSS is expected to approach completion. As such, ABYSS will be the first comprehensive spectroscopic census of young stars across the entire sky.

\section{Summary}

We present the selection strategy of young stars across the entire sky targeted by SDSS-V. The target catalogues are released publicly as a part of data release 18. In total, this selection has resulted in a sample of $\sim$200,000 sources, down to the limiting magnitude of $H<13$ mag or $G_{RP}<15.5$ mag, encompassing both diffuse associations as well as compact massive complexes of young stars across a range of distances, from the solar neighbourhood to the inner Galaxy. In future years, either optical or near-infrared spectra (or both) are expected to be available for most of these sources.

The selection strategy for ABYSS is complex, and relies on a variety of different tracers of youth, in an attempt to create as complete, as homogeneous, and as clean sample as possible, across all masses and ages younger than $\sim$30 Myr. A preliminary examination of the data shows that the vast majority of the sources observed to-date exhibit spectroscopic signatures of youth, although some fraction of contamination from more evolved sources (main sequence or red giants) is present across all cartons. Future studies employing larger data sets will precisely quantify the contamination level and allow to select cleaner samples. The development of  dedicated pipelines to derive accurate stellar parameters is underway.

The previous iterations of SDSS have produced a spectroscopic census of young stars across several selected star forming regions. These data have been instrumental in understanding the star formation history and the three-dimensional kinematics of star forming regions as a whole \citep[e.g.,][]{foster2015,da-rio2016,stutz2016,galli2019, kounkel2022b}, as well as properties of individual stars, such as accretion \citep{campbell2023}, multiplicity \citep{kounkel2019},  evolutionary properties \citep[e.g.,][]{serna2021,cao2022}, and stellar parameters  \citep[e.g.,][]{roman-lopes2019,ramirez-preciado2020}. Similarly to SDSS, other spectroscopic surveys have targeted  nearby star forming regions, such as GALAH \citep{kos2021}, or with Gaia-ESO \citep[e.g.,][]{sacco2015,bouvier2016}. Taken together however, these efforts amounted to only $\sim$10,000 young stars across a very limited footprint on the sky that generally could not encompass even a given extended population in full.
ABYSS will expand the total spectroscopic census of young stars by more than an order of magnitude, without the stringent spatial restrictions that were necessary in the past and it will substantially increase our ability to probe the recent epoch of star formation in the Galaxy as a whole. Finally, while sophisticated, the homogeneous selection function described in this study is a significant improvement on previous efforts that were specifically tailored to individual star forming regions \citep{roman-zuniga2023}, as it enables a more direct comparison between them.

In the next few years this survey will be complemented by the 4MOST Survey of Young Stars (4SYS), and we expect future \textit{Gaia} data releases to improve the processing of the RVS spectra of young stars. However, ABYSS is the first major spectroscopic survey to focus on young stars across the sky and to provide accurate stellar parameters for large, statistical samples. With yearly data releases (DR19 onward) yielding spectra and stellar parameters, it is our hope that these data will be of value to the community. 

\software{TOPCAT \citep{topcat}, PyXCSAO, APOGEE Net}

\acknowledgments
AS gratefully acknowledges support by the Fondecyt Regular (project
code 1220610), and ANID BASAL projects ACE210002 and FB210003. C.R-Z acknowledges support from projects  CONACYT CB2018 A1S-9754, Mexico and UNAM DGAPA PAPIIT IN112620, Mexico. The research leading to these results has (partially) received funding from the KU~Leuven Research Council (grant C16/18/005: PARADISE) and from the BELgian federal Science Policy Office (BELSPO) through PRODEX grant PLATO.
K.P.R. acknowledges support from ANID FONDECYT Iniciaci\'on 11201161. A.B. acknowledges partial funding by the Deutsche Forschungsgemeinschaft Excellence Strategy - EXC 2094 - 390783311 and the ANID BASAL project FB210003. RLV acknowledges support from CONACYT through a postdoctoral fellowship within the program ‘Estancias Posdoctorales por México’.  
B.R-A acknowledges funding support from FONDECYT Iniciación grant 11181295 and ANID Basal project FB210003.

Funding for the Sloan Digital Sky Survey V has been provided by the Alfred P. Sloan Foundation, the Heising-Simons Foundation, the National Science Foundation, and the Participating Institutions. SDSS acknowledges support and resources from the Center for High-Performance Computing at the University of Utah. The SDSS web site is \url{www.sdss5.org}.

SDSS is managed by the Astrophysical Research Consortium for the Participating Institutions of the SDSS Collaboration, including the Carnegie Institution for Science, Chilean National Time Allocation Committee (CNTAC) ratified researchers, the Gotham Participation Group, Harvard University, Heidelberg University, The Johns Hopkins University, L'Ecole polytechnique f\'{e}d\'{e}rale de Lausanne (EPFL), Leibniz-Institut f\"{u}r Astrophysik Potsdam (AIP), Max-Planck-Institut f\"{u}r Astronomie (MPIA Heidelberg), Max-Planck-Institut f\"{u}r Extraterrestrische Physik (MPE), Nanjing University, National Astronomical Observatories of China (NAOC), New Mexico State University, The Ohio State University, Pennsylvania State University, Smithsonian Astrophysical Observatory, Space Telescope Science Institute (STScI), the Stellar Astrophysics Participation Group, Universidad Nacional Aut\'{o}noma de M\'{e}xico, University of Arizona, University of Colorado Boulder, University of Illinois at Urbana-Champaign, University of Toronto, University of Utah, University of Virginia, Yale University, and Yunnan University.

This work has made use of data from the European Space Agency (ESA)
mission {\it Gaia} (\url{https://www.cosmos.esa.int/gaia}), processed by
the {\it Gaia} Data Processing and Analysis Consortium (DPAC,
\url{https://www.cosmos.esa.int/web/gaia/dpac/consortium}). Funding
for the DPAC has been provided by national institutions, in particular
the institutions participating in the {\it Gaia} Multilateral Agreement.

\bibliographystyle{aasjournal.bst}
\bibliography{ms.bbl}

\end{document}